\documentclass[pdflatex,sn-mathphys-num]{sn-jnl}% Math and Physical Sciences Numbered Reference Style
\usepackage{graphicx}%
\usepackage{multirow}%
\usepackage{amsmath,amssymb,amsfonts}%
\usepackage{amsthm}%
\usepackage{mathrsfs}%
\usepackage[title]{appendix}%
\usepackage{xcolor}%
\usepackage{textcomp}%
\usepackage{manyfoot}%
\usepackage{booktabs}%
\usepackage{algorithm}%
\usepackage{algorithmicx}%
\usepackage{algpseudocode}%
\usepackage{listings}%
\usepackage{bm}
\usepackage{subfig}
\usepackage{CJKutf8}
\theoremstyle{thmstyleone}%
\theoremstyle{thmstyletwo}%

\theoremstyle{thmstylethree}%

\begin{document}

\title[A highly sensitive SF$_6$-based leak test system for JUNO 3-inch PMT underwater electronics boxes]{A highly sensitive SF$_6$-based leak test system for JUNO 3-inch PMT underwater electronics boxes}


\author[1,2]{\fnm{Ziliang} \sur{Chu}}

\author[1]{\fnm{Diru} \sur{Wu}}

\author*[1]{\fnm{Miao} \sur{He}}\email{hem@ihep.ac.cn}

\author[1]{\fnm{Jilei} \sur{Xu}}

\author[1]{\fnm{Xiaoping} \sur{Jing}}

\author[1]{\fnm{Jian} \sur{Wang}}

\affil[1]{\orgname{Institute of High Energy Physics}, \orgaddress{\city{Beijing}, \postcode{100049}, \country{China}}}

\affil[2]{\orgname{University of Chinese Academy of Science}, \orgaddress{\city{Beijing}, \postcode{100049}, \country{China}}}


\abstract{A total of 25600 3-inch photomultiplier tubes (PMTs), along with their corresponding frontend electronics, have been installed at the Jiangmen Underground Neutrino Observatory (JUNO). These electronics are housed in 200 stainless steel boxes that operate underwater. To verify the sealing integrity of the underwater boxes following integration, we developed an SF$_6$-based leak test system, opting against the typical helium-based system due to helium's ability to penetrate the PMT glass. After a few hours of accumulating leaking SF$_6$ from the underwater boxes, a leak rate detection limit of $2.3\times{10}^{-9}$~Pa$\cdot$m$^3$/s in terms of SF$_6$ was achieved, corresponding to $1.65\times{10}^{-8}$~Pa$\cdot$ m$^3$/s helium equivalent. This meets the sensitivity requirement of 1$\times$10$^{-7}$~Pa$\cdot$m$^3$/s. This system was critical in identifying and replacing a few cases of leaking underwater boxes before installation.}




\maketitle

\section{Introduction}\label{sec.intro}

Jiangmen Underground Neutrino Observatory (JUNO)~\cite{JUNO:2022hxd}
%,JUNO:2024fdc,JUNO:2023ete,JUNO:2020xtj,JUNO:2022mxj,JUNO:2022hlz,JUNO:2023zty,JUNO:2024jaw,JUNO:2022jkf}
is a neutrino experiment in Guangdong province in China. A 35.4-m diameter acrylic sphere filled with 20~kton organic liquid scintillator is located in a 43.5-m deep water pool 700~m underground. 20012 20-inch photomultiplier tubes (PMTs) and 25600 3-inch PMTs are installed in a stainless steel structure surrounding the acrylic vessel detecting optical photons coming from particle interactions in liquid scintillator or in water. Electric signals of both types of PMTs are read out through frontend electronics sealed in stainless steel boxes placed underwater in the water pool.

Due to the smaller photocathode area, the 3-inch PMT system predominantly operates in a single photoelectron mode within the energy range of reactor antineutrinos. Consequently, it is immune to charge non-linearity and can be employed to calibrate the charge response of 20-inch PMTs~\cite{JUNO:2020xtj, Cabrera:2023dek}. It also plays significant roles in proton decay searches~\cite{JUNO:2022qgr}, enhancing event energy reconstruction~\cite{Zhang:2024okq}, improving energy resolution~\cite{JUNO:2024fdc}, and providing a semi-independent measurement of solar parameters ($\theta_{12}$ and $\Delta m^2_{21}$)~\cite{JUNO:2022mxj}.

\begin{figure}[!htb]
\centering
\includegraphics[width=0.75\textwidth]{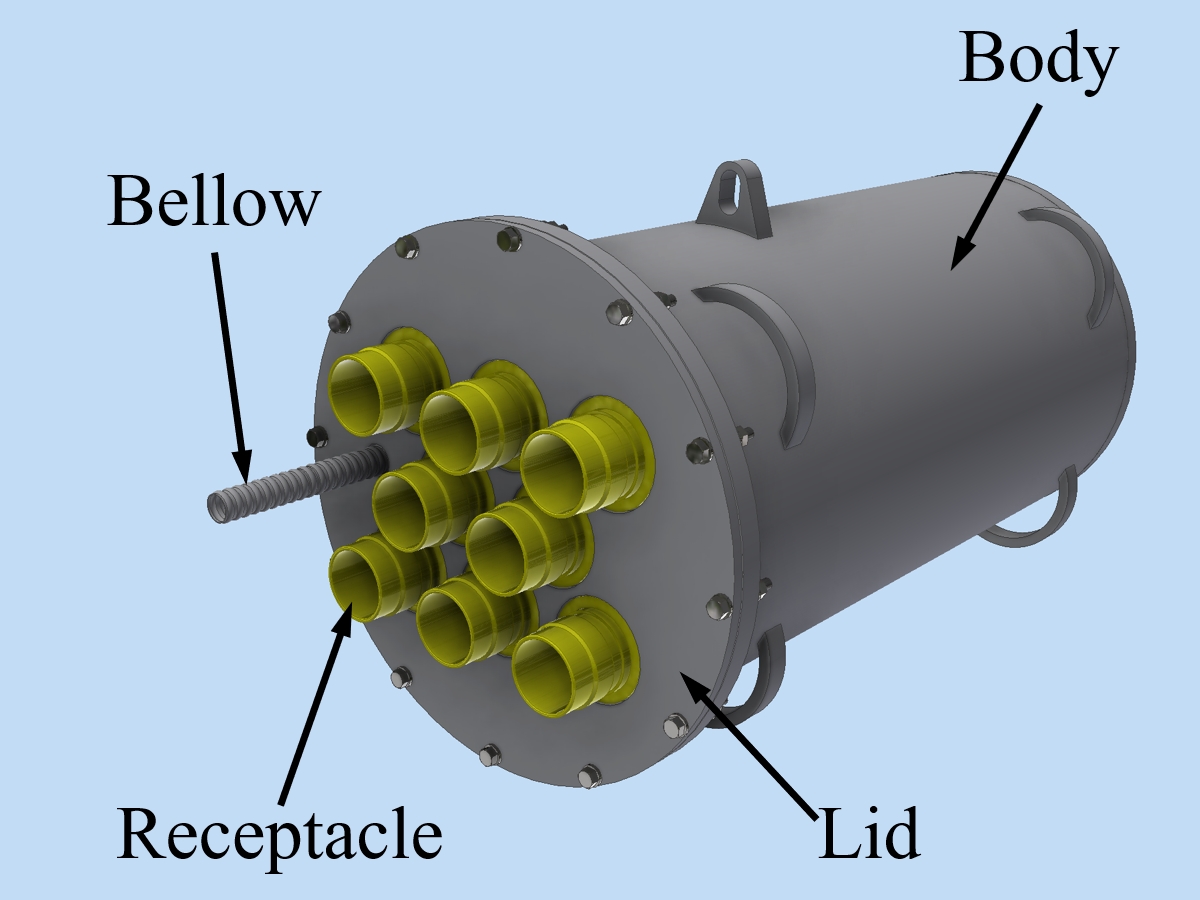}
\caption{Schematic design of the underwater electronics box. The lid has a diameter of approximately 32~cm. The body's outer diameter measures about 27~cm, with a depth of around 50~cm. The receptacles have a diameter of roughly 6~cm, while the bellow's outer diameter is about 2~cm.}
\label{fig:uwb}
\end{figure}

There are 200 underwater electronics boxes (UWBs) each serving 128 3-inch PMTs. A schematic design of the UWB is shown in Fig.~\ref{fig:uwb}. It consists of a removable lid sealed with two axial O-rings and one radial O-ring on the flange of the cylindrical body. Eight counterbores are machined on the lid, each of which holding a 16-channel receptacle sealed with two radial O-rings. The box contains four electronics boards along with heat dissipation structures. High voltage supply, signal readout, and waterproofing are realized when the receptacle is connected with the counterpart plug attached with 16 PMTs. A stainless steel bellow with a length of tens of meters is welded on the lid to protect cables communicating between front-end electronics underwater and back-end electronics on the surface.

UWBs are deployed underwater at a maximum depth of 40~m, and were planned to work for 20 years. Ensuring the long-term waterproof performance of UWBs is critical. The long-term reliability of the sealing for each O-ring, located between the lid and the body as well as between the receptacle and the lid, was assessed using several samples submerged in a water tank. To test the seal between the lid and the body, samples were subjected to a pressure of 0.8 MPa and a temperature exceeding 60~$^\circ$C for a duration of 135 days. For the seal between the lid and the receptacles, samples underwent a pressure of 0.75 MPa for 262 hours. Additionally, an entire UWB was exposed to a pressure exceeding 0.4 MPa and a temperature reaching 88.7~$^\circ$C for 8 hours. No leaks were detected during these tests.

During the mass production of UWBs, machining precision of all the sealing surface, including flanges, grooves, and counterbores, were verified through leak tests with a Helium spectrometer. A leak test for each UWB is still required when all electronics components are assembled and the box is sealed before transporting to the underground experiment hall. The test was supposed to catch possible misassembly of O-rings or defections of UWBs during transportation or electronics integration. Testing with gas is preferred compared to with water since gas is nondestructive to the UWB in case of any leak and it has higher sensitivity.

In this paper, we present the design and application of a highly sensitive leak test system utilizing SF$_6$ gas. In Sec.~\ref{sec.model}, we explore a leak model and derive the requirements for the leak test. Sec.~\ref{sec.methods} introduces various leak test methods and explains the rationale for selecting SF$_6$ as the trace gas. Both the direct measurement and the accumulation method based on SF$_6$ gas are introduced together with their sensitivities. A specialized leak test system for the JUNO 3-inch PMT underwater electronics boxes is detailed in Sec.~\ref{sec.system}, with its application to JUNO covered in Sec.~\ref{sec.ApplicationToJUNO}. In Sec.~\ref{sec.cascade} we examine the cascade leak and its impact on the sensitivity of the leak test. The summary is provided in Sec.~\ref{sec.summary}.

\section{Leak models}
\label{sec.model}

Leaks can be conceptualized as capillary tubes with intricate shapes. The flow of gas within these capillary tubes can be categorized into four primary states: turbulent flow, viscous flow, viscous-molecular flow, and molecular flow. Additionally, there may be intermediate states between these main categories. Under the specific conditions of interest, the gas flow is too minimal to exhibit turbulent behavior. 

The flow regime can be distinguished by the product of the average pressure and the diameter of the tube as shown in Table~\ref{tab:comp}~\cite{VaccumDesign}:

\begin{table}[!h]
  \centering
  \begin{tabular}{cc}
    \hline
    P$\cdot$d $\geq$ 0.67~Pa$\cdot$m & viscous flow \\
    0.02~Pa$ \cdot$m $\leq$ P$\cdot$d $<$ 0.67~Pa$ \cdot$m & viscous-molecular flow \\
    % P$\cdot$d $\geq$ 0.02~Pa$ \cdot$m and P$\cdot$d $<$ 0.67~Pa$\cdot$m & viscous-molecular flow \\
    P$\cdot$d $<$ 0.02~Pa$\cdot$m & molecular flow \\
    \hline
	\end{tabular}
	\caption{Category of gas flow.}
	\label{tab:comp}
\end{table}

\subsection{Leak rate}

For viscous flow, the volume flow rate for both liquid and gas through a circular capillary tube can be calculated using the Hagen-Poiseuille function
\begin{equation}\label{eq:leakwater}
Q_{V}=\frac{\pi d^4\Delta P}{128\mu L},
\end{equation}
where $\mu$ is the viscosity coefficient in N$\cdot$s/m$^2$, $\Delta P$ is the pressure difference in Pa, $d$ and $L$ are the diameter and length of the tube in m, respectively. This function applies to both liquids and gases. Considering the large compression ratio, it is inconvenient to use the volume flow rate defied as Eq.~\ref{eq:leakwater} for gas. It is more reasonable to define the gas flow rate as
\begin{equation}\label{eq:leakgas_viscous}
Q_{\rm gas}=\overline{P}Q_{V}=\frac{\pi d^4\Delta P}{128\mu L}\times \overline{P}
=\frac{1}{2}\frac{\pi d^4}{128\mu L}(P^2_{H}-P^2_{L}),
\end{equation}
where $P_H$ and $P_L$ are the higher and lower pressure on one side of the tube, respectively. $\overline{P}$ is the average pressure in the tube. 

For gas in viscous-molecular flow, leak rate can be calculated by:
\begin{equation}\label{eq:leakgas_viscous-molecular}
\begin{split}
Q_{\rm gas}=\frac{\pi d^4\overline P}{128\mu L}\times\Delta{P} 
+ \frac{1}{6}\sqrt{\frac{2\pi RT}{M}}\frac{d^{3}}{L}\frac{1+\sqrt{\frac{M}{RT}}\cdot\frac{d\overline{P}}{\mu}}{1+1.24\sqrt{\frac{M}{RT}}\cdot\frac{d\overline{P}}{\mu}}\times\Delta{P}
\end{split}
\end{equation}
where $R$ is molar gas constant 8.3143~J/(K$\cdot$mol), $T$ is temperature in K, $M$ is molar mass in kg/mol.

For gas in molecular flow, leak rate can be calculated by:
\begin{equation}\label{eq:leakgas_molecular}
\begin{split}
Q_{\rm gas}=\frac{1}{6}\sqrt{\frac{2\pi RT}{M}}\frac{d^{3}}{L}\Delta{P}
\end{split}
\end{equation}

\subsection{Requirement of the leak test}
\label{Target}

If a minor leak occurs in the UWB, water can enter and evaporate. During the electronics integration process, the laboratory's maximum relative humidity was approximately 66\% at a temperature of around 23°C, equating to an absolute humidity of 1855.3~Pa~\cite{GB/T11605-2005}. In the JUNO water pool, the water temperature is 21°C, resulting in a saturated water vapor pressure of 2488.1~Pa. To prevent reaching the dew point, the permissible water vapor infiltration into the UWBs is 19.0~Pa$\cdot$m$^3$, given the approximate volume of 0.03~m$^3$ within the UWB.  If we aim to keep the UWB free from liquid water over a decade, considering the deepest UWB positioned about 40~m underwater as a worst-case scenario, the maximum permissible leak rate is 4.4$\times$ 10$^{-13}$~m$^3$/s for liquid water, or 6.0$\times$10$^{-8}$~Pa$\cdot$m$^3$/s for water vapor. According to Eq.~\ref{eq:leakwater} and Eq.~\ref{eq:leakgas_viscous}, this can be translated into 7.4$\times$10$^{-7}$~Pa$\cdot$m$^3$/s for helium, or 2.3$\times$10$^{-7}$~Pa$\cdot$m$^3$/s for SF$_6$. Therefore, the requirement for the SF$_6$-based leak test is defined as 1$\times$10$^{-7}$~Pa$\cdot$m$^3$/s. In practice, the UWB is linked to the surface via the bellow, and the bellow's exit is safeguarded with nitrogen gas. This setup aids in water evaporation within the UWB, thereby mitigating the water leak requirements as well as the demands of the leak test.

%water is 4.4 $\times$ 10$^{-13}$~m$^3$/s, equivalent to an allowable leak hole diameter of 2.6$\times$10$^{-6}$~m. Under these conditions, the maximum permissible leak rates for helium mass spectrometer leak detection systems and the SF$_6$ leak detection system, which is discussed in Sec.~\ref{sec.system}, were 7.4$\times$10$^{-7}$~Pa$\cdot$m$^3$/s and 2.3$\times$10$^{-7}$~Pa$\cdot$m$^3$/s, respectively.

%UWBs were placed in the water pool for ten years, with the deepest placement at approximately 40~m. The open ends of bellows were sealed with nitrogen gas. As shown in Fig.~\ref{fig:uwb}, the volume of UWBs is approximately 0.03~m$^3$.

 %water vapor is 6.0$\times$10$^{-8}$~Pa$\cdot$m$^3$/s. Assuming the leaked water is in liquid form, this corresponds to a permissible 

% \section{Leak test system}
% \label{sec.system}

\section{Leak test methods}\label{sec.methods}

There are several common methods used for leak testing introduced in Ref.~\cite{Chen:2012ef}. Based on the component's internal pressure, these methods are classified into positive pressure and vacuum approaches. Both positive pressure and vacuum techniques can employ either the pressure change method or the tracer gas method. Below are some of the most widely used leak testing methods:

\begin{itemize}

    \item \textbf{Vacuum Pressure Change Test.}
    In this method, vacuum is established within the system or component, and any increase in pressure is monitored. A rise in pressure indicates the presence of a leak. The volume of UWBs is about 50~L, if pressure in UWB rise 1~Pa in several hours, the sensitivity of this method is approximately $10^{-6}-10^{-5}$~Pa$\cdot$m$^3$/s. 
    \item \textbf{Positive Pressure Decay Test.} 
    This method involves pressurizing the system or component and monitoring the pressure drop over a specified period. A significant reduction in pressure indicates the presence of a leak. Compared with vacuum pressure change test, this method needs larger pressure change because of higher initial pressure. If the pressure changes by 1\%, it needs to change by at least 1~KPa. Thus, the sensitivity of this method is approximately $0.1-1$~Pa$\cdot$m$^3$/s. 
    \item \textbf{Helium Leak Test.} 
    This method generally employs a vacuum technique and is extensively used for detecting leaks with high sensitivity. Helium serves as a tracer gas, and a mass spectrometer or helium leak detector is employed to detect any helium escaping into the system or component. Helium is often selected as the tracer gas due to its small atomic size, which facilitates easier passage through leaks, and its chemical inertness. Sensitivity of the helium leak detector we used is $5\times 10^{-11}$~Pa$\cdot$m$^3$/s~\cite{HeliumDetector}.
    \item \textbf{SF$_6$ Leak Test.} Similar to helium, SF$_6$ is another tracer gas although the sensitivity of the SF$_6$ detector is generally lower than helium. In our study, we inject SF$_6$ into components while keeping a positive pressure. Any SF$_6$ that escapes will be detected outside the system. 
\end{itemize}

%Helium was firstly considered for the UWBs leak test. However, helium has a high permeability to glass. Helium inside PMTs will increase noise and even cause damage to PMTs~\cite{INCANDELA1988237}. The permeation rate between gas and glass is mainly depend on molecular or atom size of gas~\cite{10.1063/1.1722570}. Size of SF$_6$ is large thus it's hard to permeate into glass. Since all the UWBs were integrated in the onsite laboratory where there are a large amount of PMTs, SF$_6$  was chosen as the tracer gas. The SF$_{6}$ detector we used is LF-300 type SF$_{6}$ quantitative leak detector produced by Shanghai Kstone Technology Development Co., Ltd.~\cite{LF-300}. This detector can detect concentration of SF$_{6}$ near the sniffer. The measuring range of this detector is 0.01~PPM - 10000~PPM, with a precision of 0.01~PPM. %是否需要介绍下SF6探测器的原理？
%Add some description of the SF$_{6}$ detector.

The vacuum pressure change test or positive pressure decay test does not meet the sensitivity requirement specified in Sec.~\ref{Target}. Helium was initially considered for leak testing of the UWBs. However, its high permeability through glass can cause increased noise and potentially damage the PMTs~\cite{INCANDELA1988237}. The rate of gas permeation through glass primarily depends on the molecular or atomic size of the gas~\cite{10.1063/1.1722570}. Given that SF$_6$ has a larger molecular size, it is less likely to permeate through glass. As all UWBs are integrated within an onsite laboratory that contains numerous PMTs, SF$_6$ was chosen as the tracer gas. While commercial high-precision SF$_6$ are available, it is necessary to customize a leak test system and develop physics models to convert the measured SF$_6$ concentration into a leak rate. Systematic uncertainties of the system must be assessed. Given the large number of UWBs, it is also essential to review operational conditions and procedures to ensure the reliability of the leak test.

The SF$_6$ detector employed in our study is the LF-300 model~\cite{LF-300}, a quantitative leak detector produced by Shanghai Kstone Technology Development Co., Ltd. This detector can measure SF$_6$ concentration in the vicinity of the sniffer, with a measurement range from 0.01 parts per million (PPM) to 10,000~PPM. The output value from this detector can fluctuate by about 20\%. Additionally, the value changes when we adjust the inspiratory rate of the detector, so we consider a relative error of 50\%. The minimum value that can be read by the SF$_6$ detector is 0.01~PPM, however, taking into account the uncertainty of the measurement and the potential presence of residual SF$_6$ gas in the environment during our tests, we have established 0.02 PPM as the detection limit.

According to the design specifications for the SF$_6$ detector, we are required to use a positive pressure method, unlike the helium mass spectrometer approach. SF$_6$ should be introduced into the test object while maintaining a positive pressure. If a leak is present, the SF$_6$ inside the test object will escape at the point of the leak and can be detected by the SF$_6$ detector. There are two methods for detecting leaks using the SF$_6$ detector: direct measurement and the accumulation method.

\textbf{Direct measurement.} 
The test object is placed in the atmosphere. SF$_6$ escaping from the leak point will diffuse into the surrounding air and can be directly detected by the SF$_6$ detector positioned near the leak point. The gas diffusion equation can be expressed as:
\begin{equation}\label{eq:DiffusionFunction}
   \bm{J}=-D\nabla{}C
\end{equation}
Where $\bm{J}$ is a vector of the diffusive flux, $D$ represents the diffusion coefficient, $C$ denotes the concentration.

If the system is spherically symmetric, in a spherical coordinate system, take the leak point as the origin of coordinate system, the diffusion equation is:
\begin{equation}\label{eq:DiffusionFunction_r}
    \frac{\partial C}{\partial t}=D(\frac{\partial^{2} C}{\partial r^{2}}+\frac{2}{r}\frac{\partial C}{\partial r})
\end{equation}
Where $r$ is the diffusion distance. 

Here, the unit of concentration is expressed in mol/m$^3$. However, to facilitate comparison with the results from the SF$_6$ detector, we also utilized PPM as the unit of concentration. These two units can be converted according to the ideal gas law:
\begin{equation}\label{eq:GasStateFunction}
PV=nRT
\end{equation}
Where $P$ is pressure, $V$ is volume, $n$ is amount of substance, $R$ is gas constant 8.314~J/mol$\cdot$K. At laboratory environment, $T$ is 295~K, $P$ is 0.1~MPa. Thus, we have:
\begin{equation}
1~{\rm mol}/m^3 = 2.45\times 10^{4}~{\rm PPM}
\end{equation}

The diffusion coefficient of a binary gas can be estimated using the following equation~\cite{1966551}:
\begin{equation}
D_{AB}=\frac{0.0101T^{1.75}\sqrt{\frac{1}{M_{A}}+\frac{1}{M_{B}}}}{P[(\sum{v_{A}})^{\frac{1}{3}}+(\sum{v_{B}})^{\frac{1}{3}}]^{2}},
\end{equation}
where \\
$D_{AB}$ = binary diffusion coefficient in m$^2$/s, \\
$T$ = temperature in Kelvin, in our laboratory is 295~K, \\
$P$ = the pressure in atmospheres in Pascal, \\
$M_{A}$, $M_B$ = molar masses of gases $A$ and $B$ in $g$/mol, \\
$\sum{v_A}$, $\sum{v_B}$ = diffusion volumes in cm$^3$/mol. \\
The diffusion volumes of SF$_{6}$ and air are 69.7~cm$^3$/mol and 20.1~cm$^3$/mol~\cite{1966551}, respectively. Consequently, the diffusion coefficient for the SF$_6$-air mixture is $9.15\times{10}^{-6}{\rm m}^2/$s.

By incorporating the parameters mentioned above and utilizing numerical methods to solve Eq.~\ref{eq:DiffusionFunction}, we can determine the concentration of SF$_6$ at various times and positions. 

When the system reaches a steady state, the value of diffusive flux is $J=\frac{Q}{4\pi r^2}$ and the direction of the diffusive flux is backwards to the leak point. Thus, Eq.~\ref{eq:DiffusionFunction} can be expressed as follows at steady state:
\begin{equation}
\frac{Q}{4\pi r^2} = -D\frac{\partial C}{\partial r}
\end{equation}
$Q$ is leak rate of the leak point, and $r$ is the diffusion distance.

Solution of this function is:
\begin{equation}
C = \frac{Q}{4\pi D}r^{-1}
\end{equation}

\begin{figure}[!h]
\centering
\includegraphics[width=0.9\textwidth]{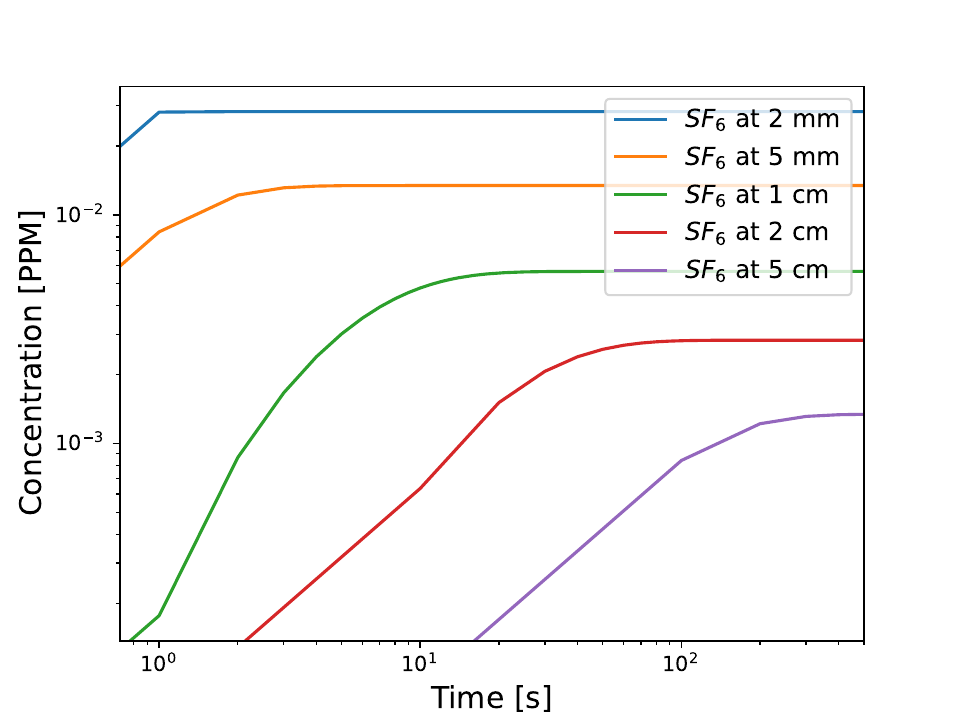}
\caption{The concentration change process of SF$_6$ at different positions when leak rate is $1\times{10}^{-9}$~Pa$\cdot$ m$^3$/s.}
\label{fig:DirectTest}
\end{figure}

For the evolution of concentration, we simulated the diffusion process distributed as Eq.~\ref{eq:DiffusionFunction} using numerical method. Fig.~\ref{fig:DirectTest} shows the simulation result with a leak rate of $1\times{10}^{-9}$~Pa$\cdot$ m$^3$/s. For positions within a distance of 5~cm from the leak point, the concentration will reach a steady state after approximately 7 minutes.

In practical application, the sniffer of the SF$_6$ detector is moved around potential leak points to identify the maximum reading. Considering the diameter of the sniffer is approximately 4~mm, we estimate the distance between the sniffer and the leak point to be 5~mm, with an uncertainty of 3~mm. This results in a sensitivity of $60^{+35}_{-36}\times{10}^{-9}$~Pa$\cdot$ m$^3$/s/PPM, which aligns with the empirical formula provided by the manufacturer of the SF$_6$ detector,  $50\times{10}^{-9}$~Pa$\cdot$ m$^3$/s/PPM. It is important to note that the diffusion of SF$_6$ can be influenced by convection; therefore, measurements should be conducted under conditions with minimal wind.

%The diameter of the sniffer of SF$_6$ detector is 4~mm. When the maximum concentration value is detected, the distance between sniffer and leak point can be estimate as 2~mm to 8~mm. Thus the sensitivity of direct test method is $24\sim95\times{10}^{-9}Pa\cdot {m}^3/s/PPM$. This result is consistent with the empirical formula provided by the manufacturer of the SF$_6$ detector which is $50\times{10}^{-9}Pa\cdot {m}^3/s/PPM$.

%In the real test environment, the distance between sniffer and leak point is hard to control. The diffusion of SF$_6$ will influenced by convection, which is complex. As a result, the sensitivity of direct measurement is not stable.

\textbf{Accumulation Method.} By sealing the leak source within a container and allowing SF$_6$ to accumulate, the concentration increases over time, thus improving detection sensitivity. This approach was selected as the primary testing method for our PMT electronics boxes. When the leak rate is minimal, it can be considered constant, as can the pressure inside the container.

If the container is not completely sealed, SF$_6$ will diffuse and escape through any leaks. Since we focused on the SF$_6$ within the container, for clarity, when discussing this method, the term ``leak rate'' refers to the leak rate of the container, while the term specific to UWB is denoted as the SF$_6$ generation rate. Similarly, ``leak point'' refers to the location of leaks in the container, whereas the term specific to UWB is identified as the SF$_6$ source location.

When the leak rate is minimal, the SF$_6$ that escapes from the container will diffuse rapidly, resulting in an external concentration of 0. According to Eq.~\ref{eq:DiffusionFunction}, the leakage of SF$_6$ from the container $Q_{\rm out}$ can be expressed as:

\begin{equation}\label{eq:SF6_out}
Q_{\rm out}=SD\frac{C}{d}=LC
\end{equation}
Where $S$ is the area of leak points, $d$ is the length of leak points, $D$ is the diffusion coefficient, $C$ is concentration near leak points in the container, and  $L=\frac{SD}{d}$ characterizes the properties of the leak point with respect to the specified gas.

If SF$_6$ in the container is uniform, the differential function of SF$_6$ concentration $C_{\rm SF_6}$ in the container can be expressed as 

\begin{equation}\label{eq:SF6_InBox}
\frac{dC_{\rm SF_6}(t)}{dt}=\frac{Q_{\rm SF_6}}{V}-\frac{LC_{\rm SF_6}(t)}{V},
\end{equation}

where $t$ is the accumulation time, $Q_{\rm SF_6}$ in Pa$\times$m$^3$/s is the generation rate of SF$_6$ gas from the source location, $LC_{\rm SF_6}(t)$ is the rate of SF$_6$ gas leaked from the container to the environmental atmosphere, and $V$ is the volume in the container. Solution of Eq.~\ref{eq:SF6_InBox} is:

\begin{equation}\label{eq:c}
C_{\rm SF_6}(t)=\frac{Q_{\rm SF_6}}{L}(1-e^{-\frac{Lt}{V}})+C_{\rm SF_6}(0)e^{-\frac{Lt}{V}},
\end{equation}

where $C_{\rm SF_6}(0)$ is the initial concentration of SF$_6$ gas in the container. Taking two extreme cases as examples. When $t \to \infty$, $C_{\rm SF_6}=\frac{Q_{\rm SF_6}}{L}$, which will be the maximum concentration in the container. On the other hand, when the leak rate of the container $L$ is negligible and the initial SF$_6$ gas concentration is 0, $C_{\rm SF_6}=\frac{Q_{\rm SF_6}}{V}t$, which is in linear correlation to the accumulation time.

\begin{figure}[!htb]
\centering
\subfloat[Initial SF$_6$ was 133~PPM uniform in the box with no SF$_6$ source in the PC box.]{\includegraphics[width=0.55\textwidth]{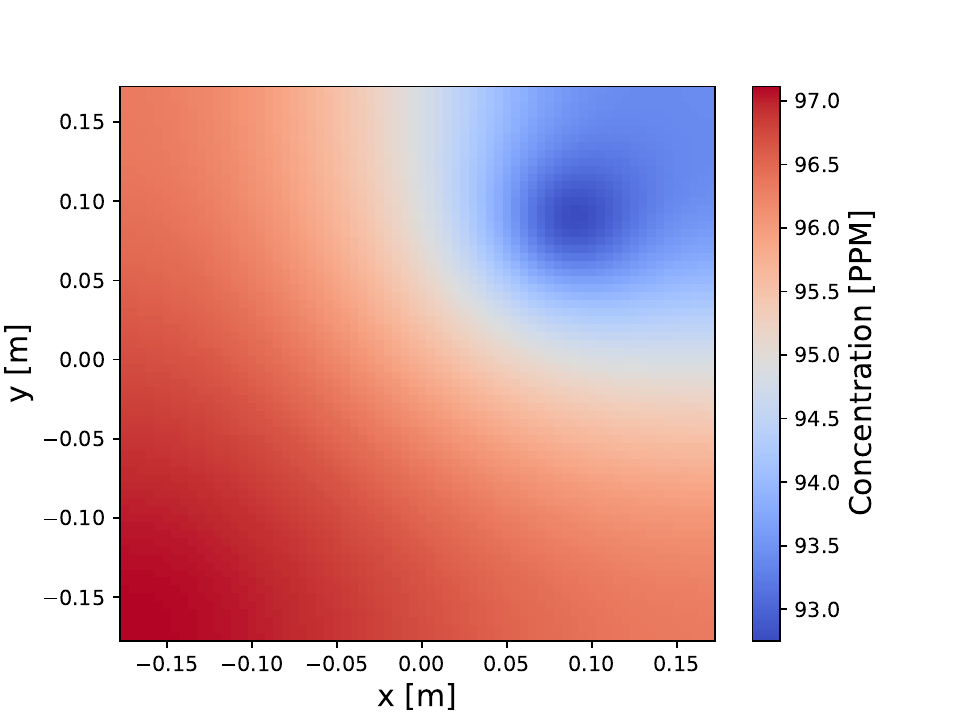}}
\\
\subfloat[Initial SF$_6$ was 0 with a SF$_6$ source on the center of PC box with generation rate is $1\times10^{-9}$~Pa$\times$m$^3$/s.]{\includegraphics[width=0.55\textwidth]{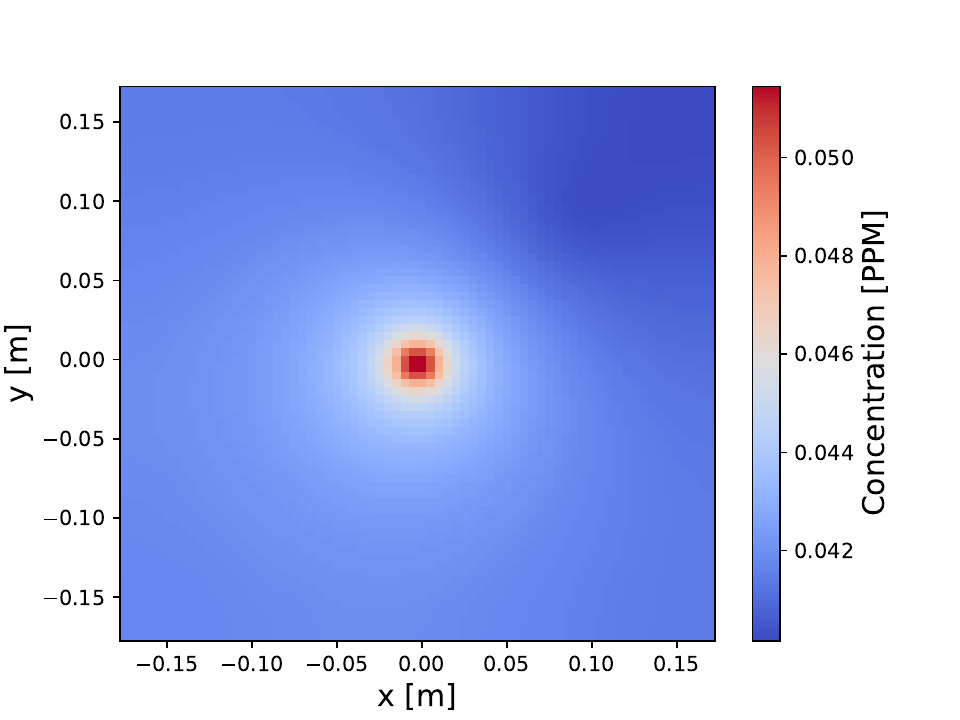}}

\caption{SF$_6$ concentration at z=0 surface after 15~hours, when leak point of PC box is at a $0.5\times0.5$~mm point on the top surface (8.75~cm, 8.75~cm, 4.5~cm) and $L=1.3\times10^{-7}$m$^3$/s.}
\label{fig:simulation}
\end{figure}

Eq.~\ref{eq:SF6_InBox} and Eq.~\ref{eq:c} are based on the assumption that the SF$_6$ concentration in the accumulation box is uniform. To evaluate the validity of this assumption, we conducted a simulation of the modeled box. According to Eq.~\ref{eq:DiffusionFunction}, we simulated a box of $35\times35\times9$~cm$^3$, which has the same dimensions and volume as the actual box described in Sec.~\ref{sec.SystemDesign}. The leak rate of the box we simulated was set to $L=1.3\times10^{-7}$~m$^3$/s. The origin is placed at the center of the box, with the x, y, and z axes pointing toward the right side, back side, and top of the box, respectively. The leak point of the box was set as a 0.5~mm $\times$ 0.5~mm spot on the top surface at coordinates (8.75 cm, 8.75 cm, 4.5 cm).  We simulated two typical cases, the calibration status and the leak detection status. In the calibration scenario, the initial concentration of SF$_6$ was set uniformly as 133~PPM throughout the box, and there was no additional SF$_6$ source present. In the leak detection scenario, the source was positioned at the center of the box with a generation rate of $1\times10^{-9}$~Pa$\times$m$^3$/s. Fig.~\ref{fig:simulation} illustrates the SF$_6$ concentration at the $z=0$ surface after 15~hours. In the calibration status, the maximum and minimum concentrations are 97.1 and 92.8~PPM, respectively. For the leak detection status, apart from locations within 5~cm of the leak point, the maximum and minimum concentrations are 0.043 and 0.040~PPM, respectively. In both scenarios, when the leak in the accumulation  box originates from a relatively small point, the maximum variation remains under 8\% except in areas very close to the SF$_6$ source. In the case of a uniform leak across the surface of the simulated box, the variation in SF$_6$ concentration within the box should be significantly smaller. These results suggest that the assumption of uniformity remains valid.%is 0.6\%.

% 
% \begin{figure}[!h]
% \subfloat[Uniform leak at the surface of the simulated box]{\includegraphics[width=0.5\hsize]{Calibration_SimPlot.pdf}}\quad
% \subfloat[Leak at a $0.5\times0.5$~mm point at the top surface (8.75~cm, 8.75~cm, 4.5~cm)]{\includegraphics[width=0.5\hsize]{Calibration_SimPlot_SmallPoint.pdf}}\quad
% \subfloat[Leak at a $0.5\times0.5$~mm point at the top surface (8.75~cm, 8.75~cm, 4.5~cm)]{\includegraphics[width=0.5\hsize]{Accumulation_SimPlot.pdf}}\quad
% \subfloat[Leak at a $0.5\times0.5$~mm point at the top surface (8.75~cm, 8.75~cm, 4.5~cm)]{\includegraphics[width=0.5\hsize]{Accumulation_SimPlot_SmallPoint.pdf}}

% \caption{SF$_6$ concentration at z=0 surface after 15~hours, when initial SF$_6$ was 133~PPM uniform in the box and $L=1.3\times10^{-7}$m$^3$/s.}
% \label{fig:simulation}
% \end{figure}
% 

\begin{figure}[!htb]
\centering
\subfloat[Minimum detectable SF$_6$ generation rate $Q_{\rm SF_6}$.]{\includegraphics[width=0.55\textwidth]{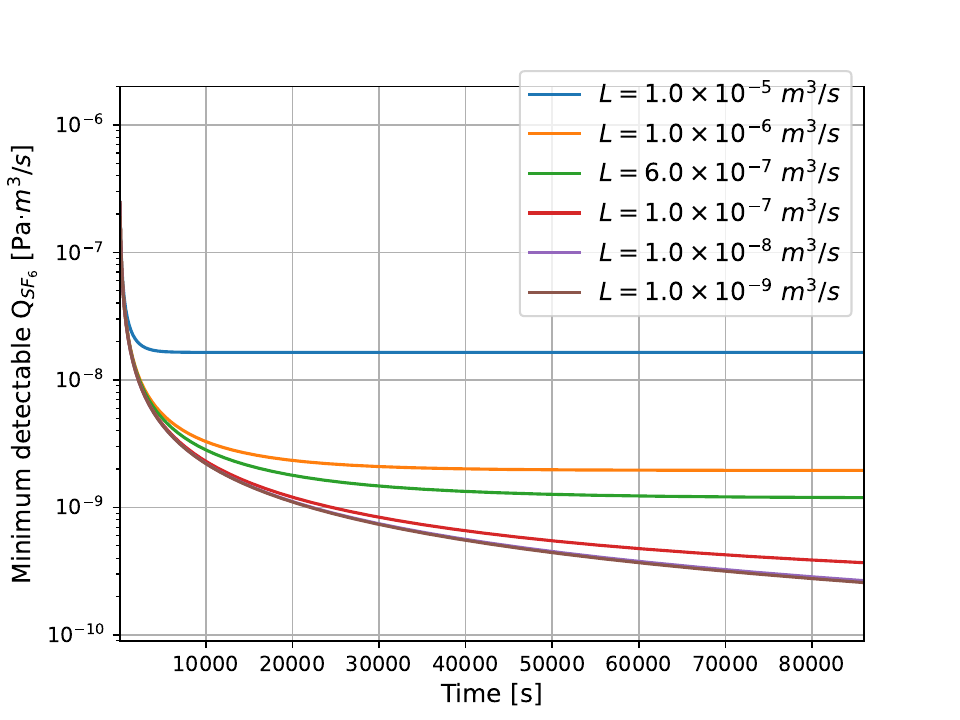}\label{fig:PCLeakrate_a}}
\\
\subfloat[Average concentrations in the accumulation box, when $Q_{\rm SF_6}$ is $5\times10^{-9}$~Pa$\cdot$m$^3$/s.]{\includegraphics[width=0.55\textwidth]{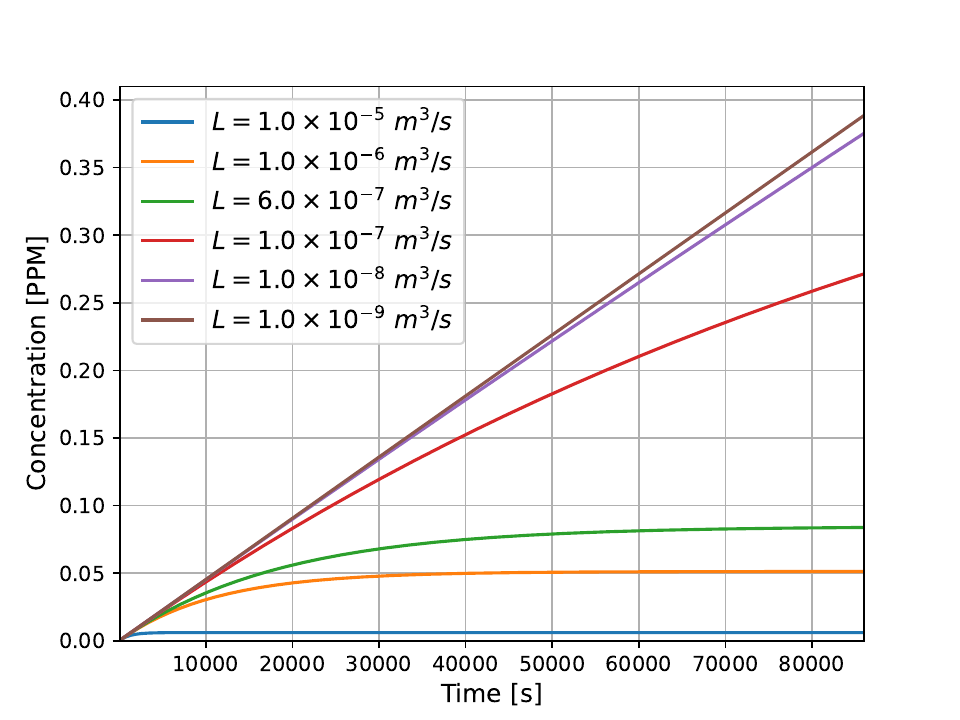}\label{fig:PCLeakrate_b}}
\caption{Sensitivity of the SF$_6$ accumulation method as a function of time with different assumptions of leak rates of accumulation box, represented with different colors.}
\label{fig:PCLeakrate}
\end{figure}

Based on this simulation, we can evaluate the sensitivity of the accumulation system at various leak rates of the accumulation box $L$. Assuming a reference SF$_6$ generation rate of $Q^{\rm ref}_{\rm SF_6}= 5\times 10^{-9}$~Pa$\times$m$^3$/s, the average concentration of SF$_6$ in the accumulation box as a function of time is presented in Fig.~\ref{fig:PCLeakrate_b}.

Taking 24 hours as the maximum testing time, when the leak rate of the accumulation box $L$ is smaller than 10$^{-8}~$m$^3$/s, the concentration is approximately linear to the accumulation time. When $L$ is smaller than 10$^{-7}~$m$^3$/s, it takes about 1.2 hours to achieve the detection limit 0.02~PPM. When $L$ is 10$^{-6}$~m$^3$/s, the minimum accumulation times is 1.6 hours. If $L$ is larger than 10$^{-5}$~m$^3$/s, SF$_6$ generated at $Q^{\rm ref}_{\rm SF_6}$ can not be detected using this method.

The SF$_6$ generation rate detection limit corresponding to 0.02~PPM concentration in the accumulation box can be thus calculated as a function of time showing in Fig.~\ref{fig:PCLeakrate_a}, presented with different assumptions of leak parameters of accumulation box. To achieve a sensitivity of 10$^{-8}$~Pa$\times$m$^3$/s SF$_6$ generation rate, the sealing of the accumulation box is crucial to control its own leak parameter $L$ smaller than $5\times 10^{-6}$~m$^3$/s. Once $L$ achieves at 10$^{-7}$~m$^3$/s, a 10$^{-9}$~Pa$\times$m$^3$ SF$_6$ generation rate can be detected within 8 hours.

Considering the concentration in the container is directly proportional to the generation rate $Q_{\rm SF_6}$, once a concentration of SF$_6$ is measured at a given time $t$, the corresponding SF$_6$ generation rate can be calculated as
\begin{equation}\label{eq:qcal}
Q_{\rm SF_6}=\frac{C^{\rm mea}(t)}{C^{\rm ref}(t)}\times Q^{\rm ref}_{\rm SF_6},
\end{equation}
where $C^{\rm mea}(t)$ is the measured concentration, $C^{\rm ref}(t)$ and $Q^{\rm ref}_{\rm SF_6}$ are the reference concentration and the reference generation rate respectively given in Fig.~\ref{fig:PCLeakrate}.

%The leakage of SF$_6$ from the container to the surrounding atmosphere is proportional to the SF$_6$ concentration within the container. Consequently, the concentration in the container should be directly proportional to the leak rate at the leak point.???

% 
% \begin{figure}[!h]
% \includegraphics[width=0.9\hsize]{PCBoxLeakrate_DetectableLeakRate.pdf}
% \caption{Sensitivity of the SF$_6$ accumulation method as a function of time with different assumptions of leak rates of accumulation box, represented with different colors. (top) Minimum detectable SF$_6$ leak rate $Q_{\rm SF_6}$. (bottom) Average concentrations in the accumulation box, when $Q_{\rm SF_6}$ is $5\times10^{-9}$~Pa$\times$m$^3$/s.}
% \label{fig:PCLeakrate}
% \end{figure}
% 

\section{Leak test system}
\label{sec.system}

\subsection{System design}\label{sec.SystemDesign}

\begin{figure*}[!htbp]
\centering
\subfloat[Schematic design of the leak test system. The underwater box (UWB) is filled with SF$_6$ gas at a pressure of 1.25 atmospheres (ATM). Any leaks occurring in the upper portion of the UWB result in the accumulation of SF$_6$ in the PC box, which can then be measured by a portable SF$_6$ detector.]{\includegraphics[width=0.7\hsize]{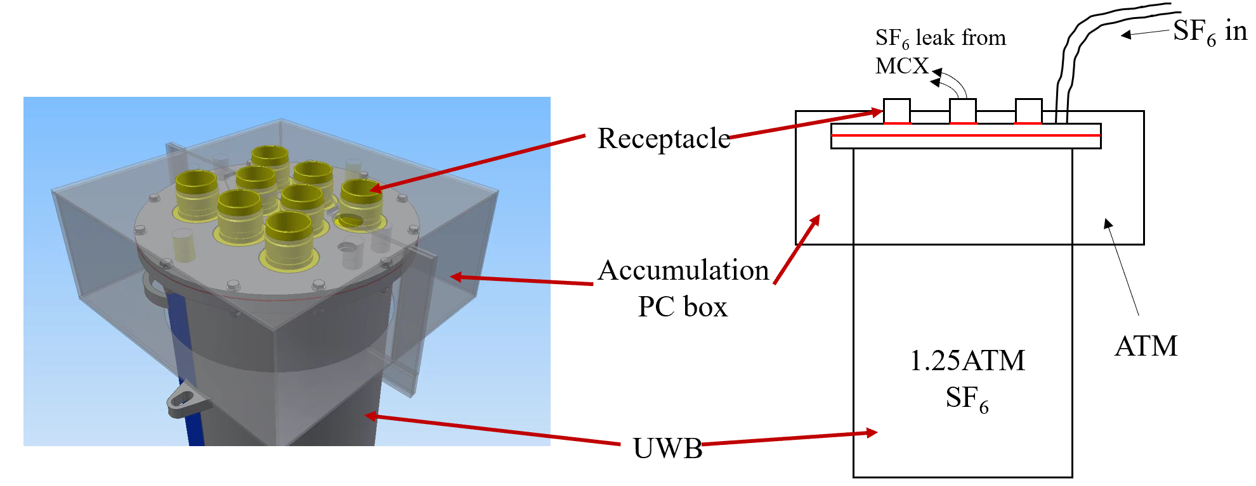}}\quad
\subfloat[Photograph of actual equipments within the leak test system.]{\includegraphics[width=0.7\hsize]{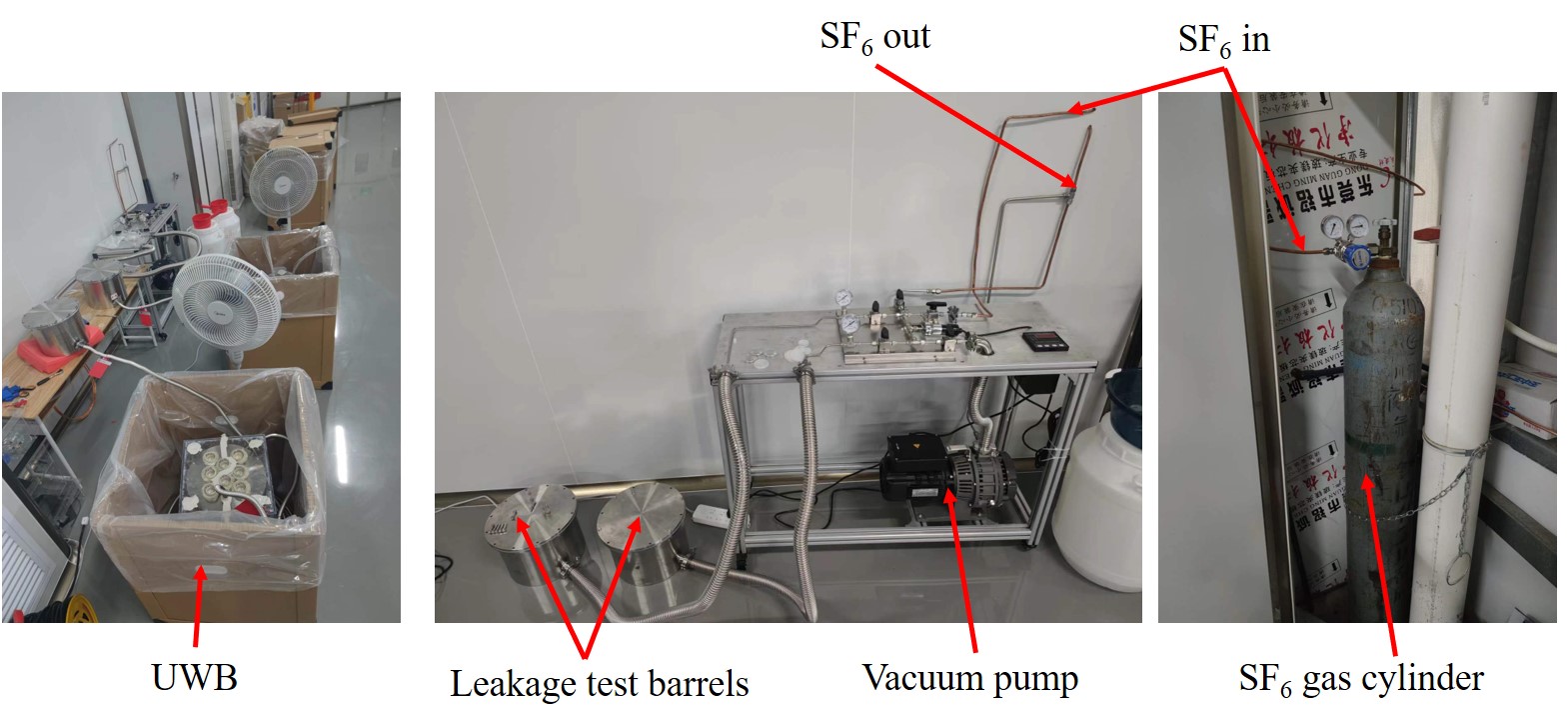}}

\caption{Schematic design and photograph of the leak test system.}
\label{fig:system}
\end{figure*}

A dedicated leak test system was designed and implemented for the JUNO 3-inch PMT underwater electronics boxes. As shown in Fig.~\ref{fig:system}, the transparent box was made of Polycarbonate (PC) and consists of two halves. There are eight holes on the top of the PC box that serve as exits for the receptacles, as they are waterproof only when connected to the plugs. Additionally, another smaller hole was used for the bellow to inject the SF$_6$ gas. In order to measure the concentration, four additional holes were drilled at the four corners of the top plate of the PC box and sealed with butyl rubber adhesive tape. When the PC box was assembled on top of the UWB, the inner volume containing air was approximately 1.1$\times10^{-2}$~m$^3$.

Eight holes for the receptacles to pass through are lined with circular rubber rings. When the receptacles are inserted through these holes, the rubber rings will be compressed, thus sealing the gaps between the PC box and the receptacles. A circle of rubber is adhered to both the edges where the two halves of the PC box come into contact and the edges where the PC box and the UWB come into contact. There are additional flat plates on both sides of the two halves of the PC box, on which three pairs of small holes are drilled. After the PC box is installed, bolts are passed through the small holes in the flat plates, so as to press tightly the rubber between the two halves of the PC box and the rubber between the PC box and the UWB. The gaps between the PC box and the UWB, as well as between the two halves of the PC box, will be additionally sealed using 2~mm-thick butyl rubber adhesive tape.

% 
% \begin{figure*}
% \includegraphics[width=0.7\hsize]{system.png}
% \caption{Schematic design of the leak test system.}
% \label{fig:system}
% \end{figure*}
% 
% 
% \begin{figure*}
% \includegraphics[width=0.7\hsize]{system.png}
% \caption{Schematic design of the leak test system.}
% \label{fig:system}
% \end{figure*}
% 
The bellow was connected to an SF$_6$ supplier and a pump via a bypass valve. Two stainless-steel barrels are used to gather the network cables that extend beyond the length of the bellows. The gap between the lids and flanges of the stainless-steel barrels is sealed with rubber gaskets and tightened by 12 sets of bolts to ensure there is no air leakage. When the test commenced, the air inside the UWB was evacuated until the residual pressure fell below 200~Pa, which typically took less than two minutes. After this, SF$_6$ gas was filled until the pressure reached 0.025$-$0.030~MPa higher than atmospheric pressure. Therefore, the residual air in the UWB is less than 1\%. The valve was then closed, allowing the SF$_6$ gas to accumulate in the PC box in the event of any leaks.

% 这一段，把范围改成误差的形式，取中心值
% Sealing of the PC box itself was calibrated by injecting some SF$_6$ gas through one of the holes. The PC box was shaken together with the UWB to accelerate SF$_6$ diffusion. After that, the SF$_6$ concentration was measured though holes as the initial status. The holes were sealed again and the system was kept for several hours and measured a few times during this period. The results were fitted by Eq.~\ref{eq:c} to get the leak rate of the box. The calibration was done five times with different initial concentration form several~PPM to about 100~PPM. Results are summarized in Fig.~\ref{fig:calibration}. The PC boxes were sealed manually, as a result, leak rate of PC boxes changed at different times. Result shows $L/V=2.9\pm2.3\times10^{-5}~$s$^{-1}$. Take $V=1.1\times10^{-2}~$m$^3$, the leak rate of PC boxes is $L=3.2\pm2.5\times10^{-7}~$m$^3/s$. 

\begin{figure}[!htb]
\centering
\includegraphics[width=0.75\textwidth]{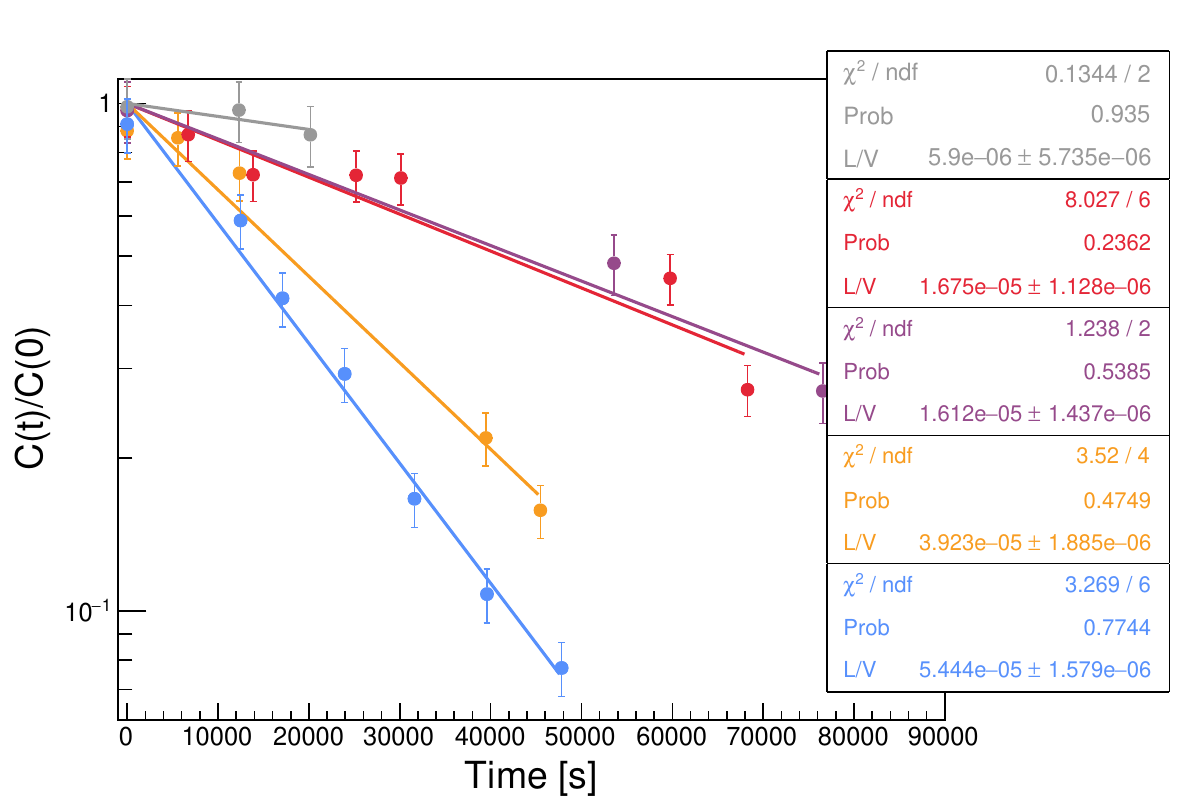}
\caption{Relative SF$_6$ concentration over time to calibrate the leak parameter $L/V$ of the accumulation box. Colors represent different calibration experiments where each time the box was cleaned, filled with SF$_6$, and sealed again. The data points are fitted with Eq.~\ref{eq:c}.}
\label{fig:calibration}
\end{figure}

Before starting the testing campaign of the PMT electronic boxes, the sealing integrity of the PC box was calibrated by injecting SF$_6$ gas through one of the holes. The PC box was then shaken together with the UWB to promote SF$_6$ diffusion. Subsequently, the SF$_6$ concentration was measured through the holes to establish the initial status. The holes were resealed, and the system was maintained for several hours, with measurements taken multiple times during this period. The results were fitted using Eq.~\ref{eq:c} to determine the leak parameter $L/V$ of the box. This calibration procedure was repeated five times with varying initial SF$_6$ concentrations, ranging from a few PPM to approximately 100~PPM. Between each trial, the PC box was reopened and resealed. The results are summarized in Fig.~\ref{fig:calibration}. 

A variation of $L/V$ was observed ranging from 0.6$\times10^{-5}$~s$^{-1}$ to 5.2$\times10^{-5}$~s$^{-1}$. This corresponds to a leak rate of the PC box $L$ from 0.7$\times10^{-7}$~m$^3$/s to 5.7$\times10^{-7}$~m$^3$/s when $V=1.1\times10^{-2}$~m$^3$. The variation was attributed to manual sealing procedures during this calibration or when testing different UWBs in the future. Therefore, we conservatively estimate the leak rate of the PC box as lying between $10^{-7}$~m$^3$/s and $10^{-6}$~m$^3$/s or expressed as $L=5.3\pm4.7\times10^{-7}$~m$^3/s$. 

% \textcolor{red}{Taking the maximum $L=10^{-6}$~m$^3$/s, the detection limit of this system can be determined using the orange curve in Fig.~\ref{fig:PCLeakrate}. The SF$_6$ concentration becomes stable after approximate 5 hours of accumulation. Taking the minimum accumulation time of 5.6 hours, as described in Sec.\ref{sec.ApplicationToJUNO}, the minimum detectable SF$_6$ leak rate is $2.4\times{10}^{-9}$~Pa$\cdot$ m$^3$/s, and it is reduced to $2.0\times{10}^{-9}$~Pa$\cdot$ m$^3$/s according to Fig.~\ref{eq:c} after accumulating for infinite time. If we can improve the leak rate of the PC box to $L=10^{-7}$~m$^3$/s, extending the accumulation time could significantly improve the detection limit, for example, achieving $3.8\times{10}^{-10}$~Pa$\cdot$ m$^3$/s  after accumulating for 24 hours.}

With a maximum leak rate of $L=10^{-6}$~m$^3$/s, the system's detection limit can be evaluated using the orange curve in Fig.~\ref{fig:PCLeakrate}. The SF$_6$ concentration stabilizes after roughly 5 hours of accumulation. Considering a minimum accumulation time of 5.6 hours during mass leak testing of JUNO UWBs, as reported in Sec.\ref{sec.ApplicationToJUNO}, the smallest detectable SF$_6$ leak rate is $2.4\times{10}^{-9}$~Pa$\cdot$ m$^3$/s. This rate decreases to $2.0\times{10}^{-9}$~Pa$\cdot$ m$^3$/s after infinite accumulation time, as illustrated in Eq.~\ref{eq:c}. If we enhance the leak rate of the PC box to $L=10^{-7}$~m$^3$/s, prolonging the accumulation time could significantly improve the detection limit, potentially reaching $3.8\times{10}^{-10}$~Pa$\cdot$ m$^3$/s after 24 hours of accumulation.
 
% After about 5.6 hours, the minimum detectable SF$_6$ leak rate became stable, and further increasing the time did not significantly improve the sensitivity. It's efficient to take 5.6 hours as the accumulation time. while the minimum detectable SF$_6$ leak rate after accumulating for infinite time is  $2.0\times{10}^{-9}$~Pa$\cdot$ m$^3$/s according to Fig.~\ref{eq:c}. }

%\textcolor{red}{Furthermore, it's also worthwhile to increase accumulation time more than 5.6 hours. Because we may have lower leak rate of the PC box up to  $L=10^{-7}$~m$^3$/s. If $L=10^{-7}$~m$^3$/s, the minimum detectable SF$_6$ leak rate after accumulating for 24 hours is $3.8\times{10}^{-10}$~Pa$\cdot$ m$^3$/s.} % If there is no leak on PC box, the minimum detectable SF$_6$ leak rate after 24 hours is $2.6\times{10}^{-10}$~Pa$\cdot$ m$^3$/s. }

%\textcolor{red}{The PC box was opened and sealed every time when we tested. It's hard to control leak rate of the PC box all the times. To take a conservative estimation, and compare with the result of helium mass spectrometer discussed in Sec.~\ref{sec.comp}, we extend the maximum leak rate to 1.0$\times$10$^{-6}$~m$^3$/s, and take $L=5.3\pm4.7\times10^{-7}~m^3/s$. }

%\textcolor{red}{Combine with the simulation result shown in Fig.~\ref{fig:PCLeakrate}, after accumulation for 6 hours, if no leak was found, the maximum leak rate of UWB is $2.3\times{10}^{-9}$~Pa$\cdot$ m$^3$/s.}

\subsection{Validation of the system}
%\subsection{Comparison with the Helium system}
\label{sec.comp}
%\section{Leak models}\label{sec.model}

%\subsection{Leak through a capillary tube}
As described in Sec.~\ref{sec.ApplicationToJUNO}, during the mass integration of the 3-inch PMTs electronics, four UWBs were found to leak through SF$_6$ accumulation measurements. They were subsequently tested by the SF$_6$ direct measurements, which identified three leak points at the weld between the bellow and the lid. Two of them were relocated to a nearby building far away from PMTs, where they were measured using a Helium mass spectrometer system. The test results of the two UWBs serve to validate this system.

\subsubsection{Comparison with the direct test}

The two selected UWBs were measured twice at different times during the accumulation of SF$_6$ gas, with the results summarized in Table~\ref{tab:leak_result}. For UWB No.~4199, the SF$_6$ concentration in the PC box significantly increased from 6 hours to 15 hours, as expected. In contrast, no such increase was observed for UWB No.~4207, likely because its SF$_6$ concentration was already nearing the lower limit of the SF$_6$ detector. These results were converted to the SF$_6$ leak rate using Eq.~\ref{eq:qcal}, with uncertainties originating from both the SF$_6$ detector and the leak rates of the accumulation box $L$. The two measurements conducted at different times exhibited differences of up to a factor of 2.5, although they remained consistent within the associated uncertainties. Furthermore, both UWBs were measured directly after the removal of the PC box, which allowed for the identification of the leak points, leading to the presentation of both the concentration and the leak rate. In comparison to the accumulation method, the direct tests indicated higher leak rates, suggesting that the sniffer was nearer to the leak point than initially anticipated. Nevertheless, when all uncertainties are taken into account, both methods yielded consistent leak rates.

%For UWB No.4199, SF$_6$ concentration in the accumulation PC box was 0.17~PPM after about 22920~s and 0.6~PPM after about 57120~s. For UWB No.4207, SF$_6$ concentration in the accumulation PC box was 0.02~PPM after about 22860~s and 55560~s. SF$_6$ direct measurement were 2.0~PPM and 0.1~PPM for UWB No.4119 and No.4207.

% \begin{table}[!h]
%   \centering
%   \begin{tabular}{cccc}
%     \hline
%     Nb & First test & Second test & leak rate of SF$_6$  \\
%        & (PPM) & (PPM) & ($10^{-9}~Pa\cdot m^3/s$) \\
%     \hline
%     4207 & 0.02 (22860~s) & 0.02 (55560~s) & $0.5\sim1.6$  \\
%     4199 & 0.17 (22920~s) & 0.6 (57120~s) &  $8.8\sim35.5$ \\
%     \hline
%   \end{tabular}
%   \caption{Leak tested by the accumulation system.}
%   \label{tab:LeakBellows}
% \end{table}

\begin{table}
 \resizebox{\textwidth}{!}{
        \begin{tabular}{c|ccc|c|c}
    \hline
     UWB ID & \multicolumn{3}{c|}{SF$_6$ accumulation method} & SF$_6$ direct test & Helium\\
     & Time (s) & 22860 & 22920 & & \\
     \hline
      & Concentration (PPM) & $0.02\pm0.01$ & 
     $0.02\pm0.01$ & $0.1\pm0.05$ & \\
     4207 & Leak rate of SF$_6$ ($\times10^{-9}$~Pa~$\cdot$m$^3$/s) & $1.6^{+1.7}_{-1.1}$ & $1.1^{+1.8}_{-0.9}$ & $6^{+4.6}_{-4.7}$ & \\
     & Leak rate of Helium ($\times10^{-9}$~Pa~$\cdot$m$^3$/s) & $11.9^{+9.7}_{-7.4}$ & $8.7^{+10.6}_{-6.6}$ & $35^{+22}_{-25}$ & 30 \\
     %\multiraw{3}{*}{Nb} & Time & 22860 & 22920 & 0.1 \\
     \hline
      & Concentration (PPM) & $0.17\pm0.09$ & $0.60\pm0.30$ & $2.0\pm1.0$ & \\
     4199 & Leak rate of SF$_6$ ($\times10^{-9}$~Pa~$\cdot$m$^3$/s) & $13.3^{+25.7}_{-9.2}$ & $33.5^{+54.6}_{-26.6}$ &  $120^{+92}_{-94}$ & \\
     & Leak rate of Helium ($\times10^{-9}$~Pa~$\cdot$m$^3$/s) & $68.4^{+99.3}_{-42.6}$ & $147.8^{+185.7}_{-108.1}$ & $433^{+271}_{-313}$ & 160 \\
     \hline
    \end{tabular}}
    \caption{Test results of two leak UWBs by the SF$_6$ accumulation method, the SF$_6$ direct test, and the Helium method.}
    \label{tab:leak_result}
\end{table}

\subsubsection{Comparison with the Helium system}

The two selected UWBs were also tested for leaks using a helium mass spectrometer system, positioned far away from PMTs. To measure the leak rate, we covered the suspected leak point with a PET bottle, sealed with butyl rubber adhesive tape, and filled the bottle with helium. After approximately 7 minutes, the helium concentration reached equilibrium, and the maximum reading from the mass spectrometer was recorded. The measured leak rate of UWBs No.4207 and No.4199 were $30\times 10^{-9}$~Pa~$\cdot$m$^3$/s and $160\times 10^{-9}$~Pa~$\cdot$m$^3$/s respectively. 

We also performed a localized helium spray test at the suspected leak point and recorded the response from the helium mass spectrometer. While this method is commonly used for leak detection (see Ref.~\cite{HeliumDetector}), it does not provide an accurate measurement of the total leakage rate. For UWB No.4207, approximately 7 minutes after helium application, the spectrometer reading peaked at $5.3\times 10^{-9}$~Pa~$\cdot$m$^3$/s. This value was documented for reference only.

Due to differences in pressure, viscosity, and molar mass between SF$_6$ and helium, their measured leak rates cannot be directly compared. To address this, we can use leak models discussed in Sec.~\ref{sec.model} to account for the distinct transport properties of each gas.

% The typical length of the leak tube is about 1mm, which is the diameter of O-rings and the thickness of bellows. 
% According to equation \ref{eq:leakgas_viscous-molecular} and table \ref{tab:comp}, the leak rate range of helium in viscous-molecular flow with $P_H = 0.1MPa$ and $P_L \approx 0$ is about $2\times 10^{-9}Pa\cdot m^{3}/s$ to $2.5\times 10^{-4}Pa\cdot m^{3}/s$, which covers typical leak rate during test. In other words, leak regime during test is viscous-molecular flow.

%Because conditions of two systems were different, leak rate results of two systems shouldn't be compared directly. Here we suspect leak points can be treated as circle capillary tubes and we can calculate leak rates using formulas described in Sec.~\ref{sec.model}.

For our tested UWBs, the typical length of the leak tube is approximately 1~mm, which corresponds to the diameter of the O-rings and the thickness of the bellows. The temperature during the test was around 295~K. Additional parameters are listed in Table~\ref{tab:Parameters}.

\begin{table}[!h]
  \centering
  \begin{tabular}{ccccc}
    \hline
    Trace gas & $P_H$ & $P_L$ & Molar mass & Viscosity coefficient \\
    & MPa & MPa & g/mol & N$\cdot$s/m$^2$ \\
    \hline
    SF$_6$ & 0.125 & 0.1 & 146.055 & $1.5\times10^{-5}$ \\ 
    He & 0.1 & 0 & 4.0 & $2.0\times10^{-5}$ \\ 
    \hline
	\end{tabular}
	\caption{Parameters to calculate leak rates of two systems~\cite{10.1063/1.1433462,LIU2025105517}.}
	\label{tab:Parameters}
\end{table}

\begin{figure}[!htb]
\centering
\includegraphics[width=0.75\textwidth]{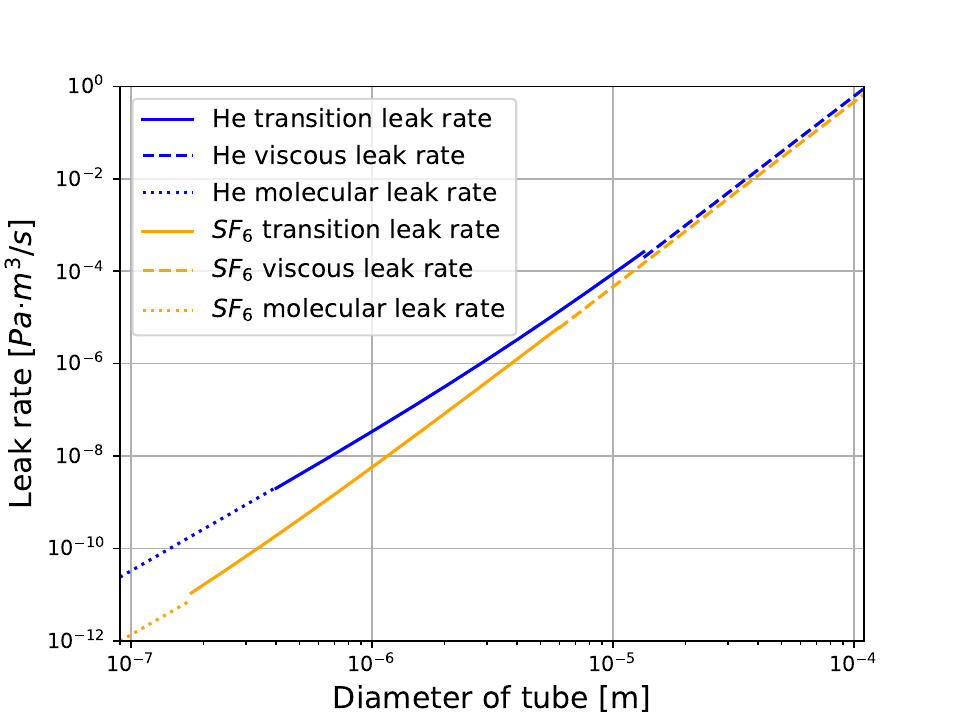}
\caption{ Leak rates of helium and SF$_6$ as a function of diameter of capillary tube.}
\label{fig:LeakRate}
\end{figure}

\begin{figure}[!htb]
\centering
\includegraphics[width=0.75\textwidth]{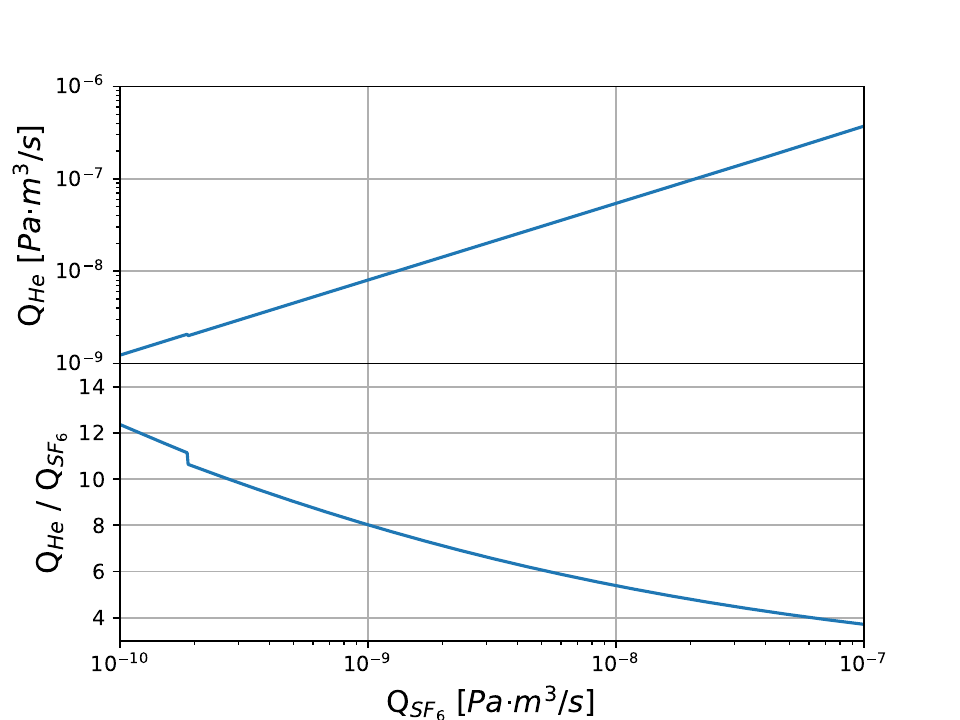}
\caption{ Comparison of leak rates of helium and SF$_6$. The small jump around 2$\times$10$^{-10}$~Pa~$\cdot$m$^3$/s originates from the transition between the molecular flow and the viscous-molecular flow.}
\label{fig:Comparison}
\end{figure}

According to Eqs.~\ref{eq:leakgas_viscous} to \ref{eq:leakgas_molecular}, the leak rates of the two systems as a function of the diameter of the capillary tube are shown in Fig.~\ref{fig:LeakRate}, while a comparison of their leak rates is illustrated in Fig.~\ref{fig:Comparison}. When the leak rate of SF$_6$ gas is 2$\times$10$^{-10}$~Pa~$\cdot$m$^3$/s, the leak rate of helium gas is ten times higher due to its greater pressure difference and lower molar mass. However, this difference diminishes as the leak rate increases.

Using the results from Fig.~\ref{fig:Comparison}, we converted the SF$_6$ leak rates to their helium-equivalent values. The normalized results for all test configurations are presented in Table~\ref{tab:leak_result}. The last column of Table~\ref{tab:leak_result} presents the measured leak rates from the helium system, with no uncertainty assigned, as this is beyond the scope of this paper. Using the helium system results as a reference, the variation in results obtained from the two SF$_6$ methods falls within a factor of 3, and they are all consistent when considering their uncertainties.

\section{Application to JUNO}
\label{sec.ApplicationToJUNO}

Leak tests using the SF$_6$ system were conducted on all 200 UWBs and a few spares following the integration of the electronics. First, the PC box was assembled onto the lid of the UWB, and the concentration of SF$_6$ within the PC box was measured. If any residual gas was detected, a fan was employed to blow air through the hole in the PC box to eliminate background interference. Next, SF$_6$ was injected into the UWB until the pressure reached 0.025$-$0.030~MPa with respect to atmospheric pressure. After a few hours of accumulation, the SF$_6$ concentration was measured through each of the four holes. If the result was greater than or equal to 0.02~PPM, it was identified as a leak. If it was 0.01~PPM, it was considered to be influenced by residual SF$_6$ in the air of the test room. In both cases, direct measurements were performed to verify the leak and to locate the leak points.

\begin{figure}[!htb]
\centering
\includegraphics[width=0.75\textwidth]{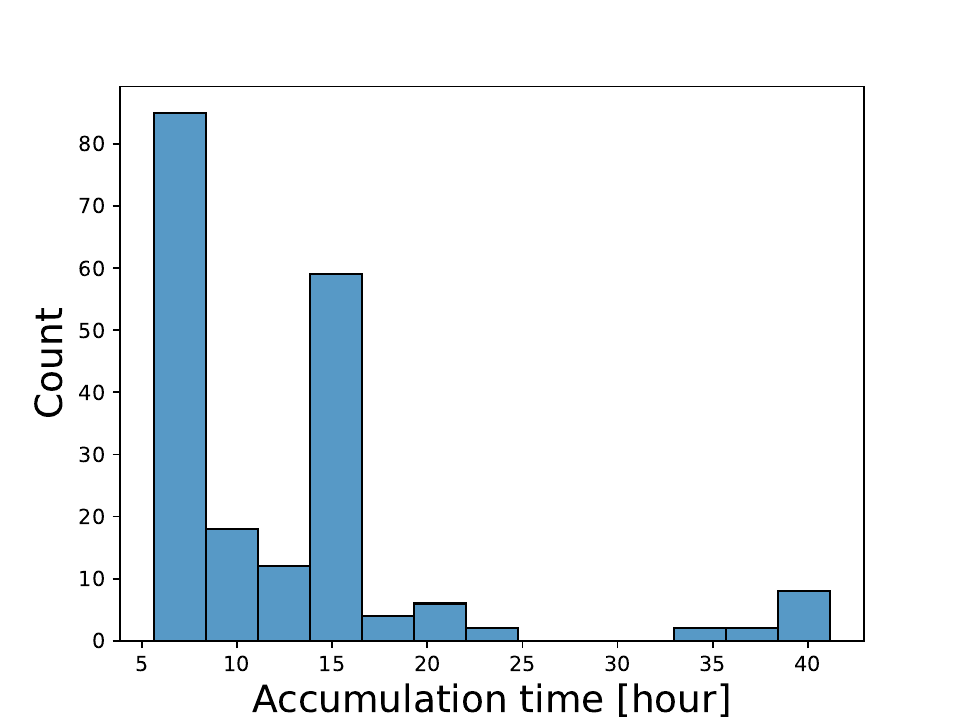}
\caption{Accumulation time distribution of all tested UWBs}
\label{fig:AccumulationTime}
\end{figure}

To meet the requirements of the JUNO schedule, three to four electronics units were integrated and tested each day. Two UWBs could be tested simultaneously with the system, and typically, two rounds of tests were conducted each day. The first group was injected with SF$_6$ in the morning, accumulated the gas for at least 5.6 hours, which is an efficient accumulation time discussed at Sec.~\ref{sec.SystemDesign}, and was tested in the afternoon. The second group was injected with SF$_6$ afterward and tested the following morning after approximately 15 hours of accumulation resulting in a higher sensitivity. Some UWBs were configured for longer accumulation times. The distribution of accumulation times is shown in Fig.~\ref{fig:AccumulationTime}. Three UWBs underwent only direct tests to keep pace with the installation schedule.

During the mass testing, a few leaking UWBs were identified. In rare cases, there was misassembly of the O-rings on the receptacles of the connectors, which resulted in significant leaks. These issues were discovered immediately after the injection of SF$_6$ when verifying residuals in the PC box. After checking and reassembling the O-rings, these UWBs were tested again.

Three UWBs exhibited leaks at the weld between the bellows and the lid, while one UWB was found to have a leak at the weld of the flange. The affected electronics were disassembled, and the leaking components were replaced. Ultimately, 203 electronics were integrated and successfully passed the SF$_6$ leak tests. Taking 5.6 hours as the minimum accumulation time, leaks in UWBs were excluded at a rate of $2.4\times{10}^{-9}$~Pa$\cdot$ m$^3$/s in terms of SF$_6$, or $1.7\times{10}^{-8}$~Pa$\cdot$ m$^3$/s in terms of helium equivalent to SF$_6$.
% Leak were then sent back to factory to repair. After repair, all UWBs passed leakage test.

\section{Cascade leak}\label{sec.cascade}

As described in Sec.~\ref{sec.intro}, there are three redundant O-rings sealing the lid with the body and two redundant O-rings sealing the receptacles with the lid. These sealing surfaces were also contained in the PC box and were thus included in the leak test. However, sensitivity of the multiple O-ring structure is significantly lower than the single O-ring case, which can be explained as cascade gaps. 

\begin{figure}[!htb]
\centering
\includegraphics[width=0.75\textwidth]{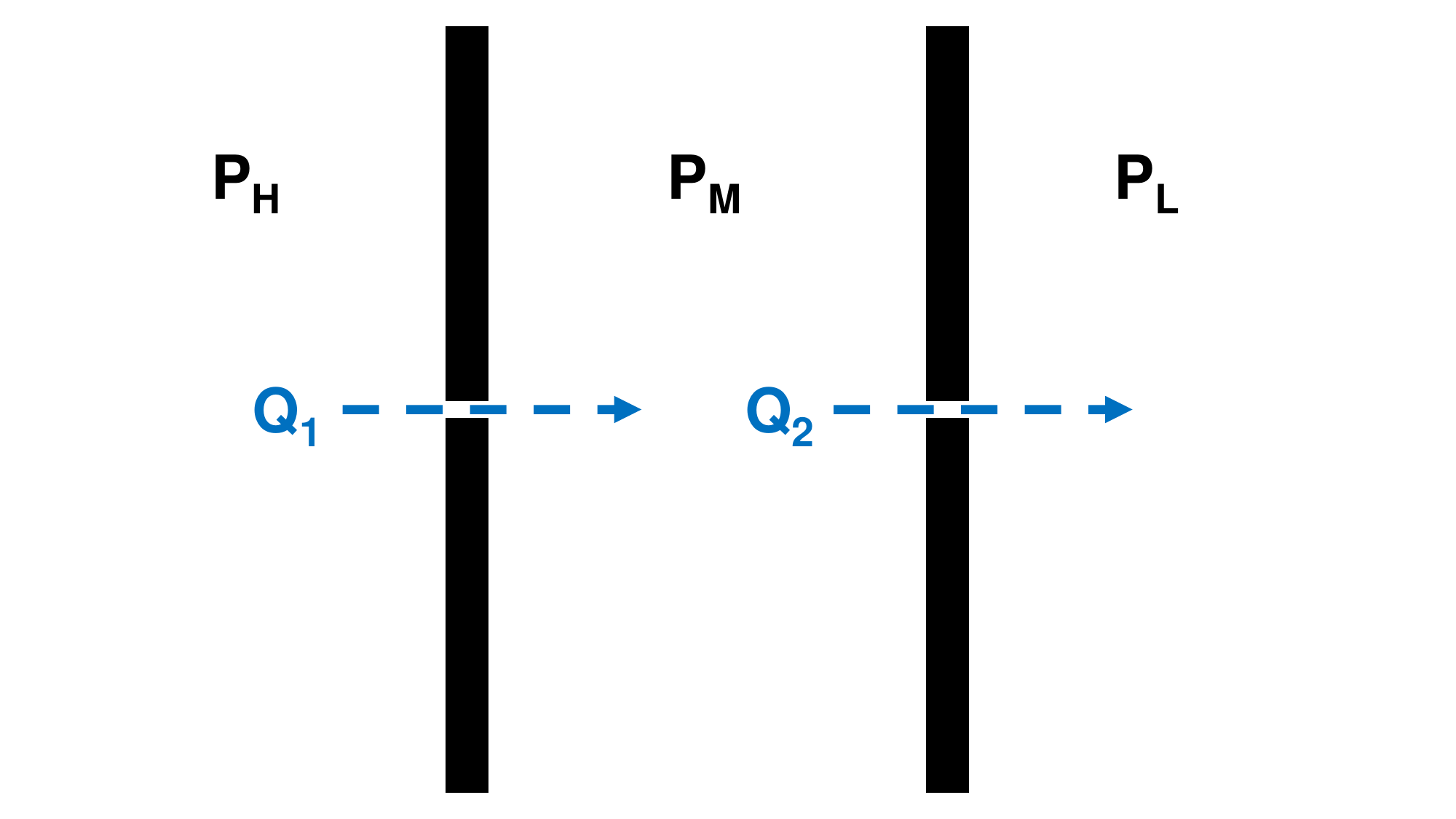}
\caption{Schematic to illustrate the cascade leak concept.}
\label{fig:casleak}
\end{figure}

As illustrated in Fig.~\ref{fig:casleak}, there are three regions with different pressures, denoted as
$P_{H}$, $P_{L}$, and $P_{M}$, where $P_{M}$ represents the medium pressure in the space between the two O-rings. Considering $P_{H}$ and $P_{L}$ are constants under normal operating conditions with a small leak, $P_{M}(t)$ is a time-dependent function that depends on the leak rate. The leak rates between each regions are defined as $Q_1$ and $Q_2$, respectively.

The leak rate of a cascade leak can be calculated based on Eqs.~\ref{eq:leakgas_viscous} to \ref{eq:leakgas_molecular}. To simplify the model, we used viscous flow as a reference. Based on Eq.~\ref{eq:leakgas_viscous}, defining $A=\dfrac{\pi d^4}{128\mu L}$ to characterise the leak tube, we can get
\begin{equation}
\begin{split}
Q_1=\frac{1}{2}A_1(P^2_{H}-P^2_{M}),\\
Q_2=\frac{1}{2}A_2(P^2_{M}-P^2_{L}).
\end{split}
\end{equation}

At a given time $t$, the differential pressure can be expressed as
\begin{equation}\label{eq:diffp}
\begin{split}
\frac{dP_{M}(t)}{dt}V=Q_1-Q_2=\frac{1}{2}(A_1P^2_{H}+A_2P^2_{L})-\frac{1}{2}(A_1+A_2)P^2_{M},
\end{split}
\end{equation}
where $V$ is the volume between two O-rings, which is assumed to be a constant. In our system, the volume between two O-rings sealing receptacles with the lid is about 0.2~mL.

When $P_{M}$ is increasing, which is the situation of positive pressure leak test method, solution of Eq.~\ref{eq:diffp} gives
\begin{equation}
\begin{split}
P_{M}(t)=\sqrt{\frac{A_1P^2_{H}+A_2P^2_{L}}{A_1+A_2}}\cdot\tanh(\frac{\sqrt{(A_1P^2_H+A_2P^2_L)(A_1+A_2)}}{2V}t)+P_{M}(0).
\end{split}
\end{equation}
When P$_{M}$ is decreasing, which is the situation of vacuum leak test method, solution of Eq.~\ref{eq:diffp} gives
\begin{equation}
\begin{split}
P_{M}(t)=\sqrt{\frac{A_1P^2_{H}+A_2P^2_{L}}{A_1+A_2}}\cdot\coth(\frac{\sqrt{(A_1P^2_H+A_2P^2_L)(A_1+A_2)}}{2V}t)+P_{M}(0).
\end{split}
\end{equation}

These two solutions have a common time constant
%\begin{equation}
$\tau=\dfrac{2V}{\sqrt{(A_1P^2_H+A_2P^2_L)(A_1+A_2)}}$,
%\end{equation}
which means that the cascade leak system will increase the time required to identify potential leak points.

When $t\rightarrow\infty$, the double O-rings system reaches an equilibrium state and $P_{M}\rightarrow\sqrt{\dfrac{A_1P^2_{H}+A_2P^2_{L}}{A_1+A_2}}$. Therefore, the total leak rate should be $Q=\dfrac{1}{2}\dfrac{A_1A_2}{A_1+A_2}(P^2_{H}-P^2_{L})$. Compare this result with Eq.~\ref{eq:leakgas_viscous}, the double O-rings system can be simplified as a single O-ring whose parameter is $A'=\dfrac{A_1A_2}{A_1+A_2}$. 

If the parameters $A_1$ and $A_2$ of the two O-rings differ significantly in order of magnitude, $A_t$ will be approximately equal to the smaller one of $A_1$ and $A_2$. This means that under these conditions, the two o-rings system can be regarded as equivalent to a single o-ring system with a smaller leak rate. Consequently, if only one O-ring is defective, we may not be able to identify the issue.
 
If $A_1=A_2=A$, then $A$' will equal $\dfrac{1}{2}A$. This means that under these conditions, the two O-rings system can be treated as a single O-ring whose leakage path is twice the length of either of the two O-rings.

%\textcolor{red}{For our system, the volume between two O-rings is about 0.2~mL. Flow regime in our system was not always viscous flow. So it's hard to get analytical results. But we can still get a numerical result based on Eqs.~\ref{eq:leakgas_viscous} to \ref{eq:leakgas_molecular}. Fig.~\ref{fig:DoubleOrings} shows the simulation of time evolution of leak rates in double O-ring sealing system. Here we assume leak points of two O-rings are the same. $Q$ is the leak rate if there is only one O-ring in the SF$_6$ leak testing system.}

\begin{figure}[!htb]
\centering
\includegraphics[width=0.75\textwidth]{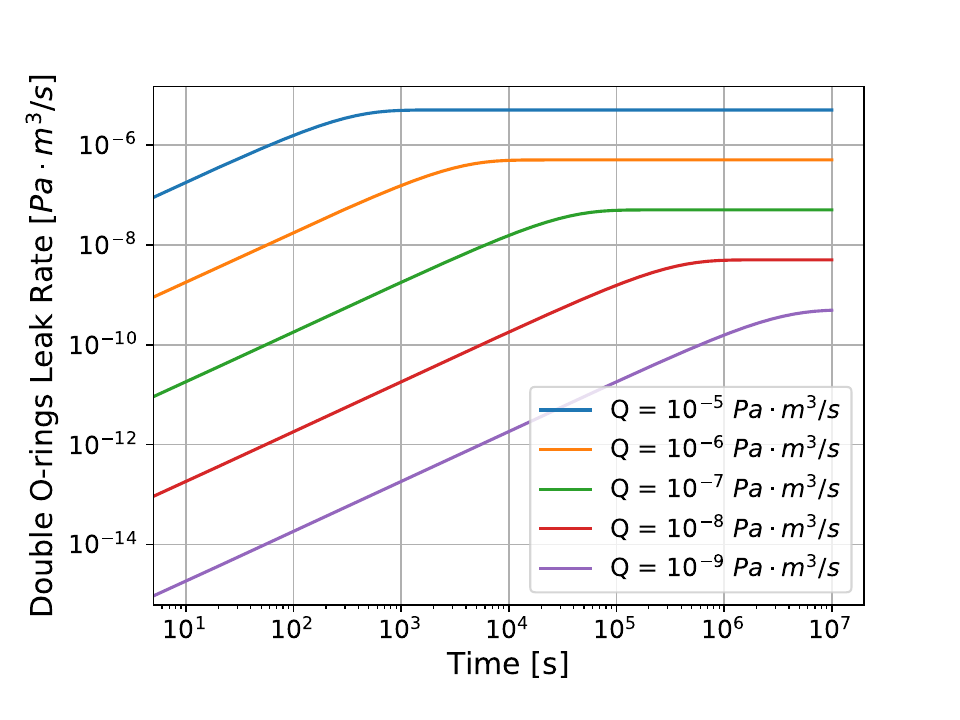}
\caption{Time evolution of the overall leak rates in the double O-rings sealing system, in our test situation described in Sec.~\ref{sec.SystemDesign}, assuming that the leak tubes of both O-rings are the same. $Q$ is the leak rate of a single O-ring.}
\label{fig:DoubleOrings}
\end{figure}

The structure of multiple O-rings generally does not consistently exhibit viscous flow, which complicates the analytical calculation of the overall leak rate. Nonetheless, approximate numerical results can be obtained using Eqs.~\ref{eq:leakgas_viscous} to \ref{eq:leakgas_molecular}. Taking the sealing of the receptacle as an example, the volume enclosed between the two O-rings is approximately 0.2~mL, and it is assumed that both O-rings have identical leak tubes. In our test situation described in Sec.~\ref{sec.SystemDesign}, the time evolution of the total leak rate is simulated and shown in  Fig.~\ref{fig:DoubleOrings}, where $Q$ denotes the leak rate of a single O-ring.

The initial overall leak rate of the double O-rings system is between 1 and 5 orders of magnitude smaller in the region we are interested in. The stabilization time is 10$^3$ seconds when $Q=10^{-5}$~Pa$\cdot$~m$^3$/s and 10$^7$ seconds when $Q=10^{-9}$~Pa$\cdot$m$^3$/s. As expected, the stabilized leak rate is $\frac{1}{2}Q$. To reach the detection limit of our system, which is $2.4\times{10}^{-9}$~Pa$\cdot$m$^3$/s, it takes 0.4~hour when $Q=10^{-7}$~Pa$\cdot$m$^3$/s and 49 hours when $Q=10^{-8}$~Pa$\cdot$m$^3$/s. During our typical testing duration of 5.6 - 42 hours, only leak rates around $Q\sim O(10^{-8})$~Pa$\cdot$m$^3$/s can be detected. This result also satisfied the sensitivity requirement specified in Sec.~\ref{Target}. Regarding  the sealing of the lid, where three O-rings are utilized, it takes a significantly longer time to reach equilibrium, and the detection limit is further reduced within a 5.6-hour time frame.

\section{Summary and discussion}
\label{sec.summary}

The leak test using SF$_6$ as the trace gas was investigated utilizing both direct measurement and accumulation methods. A gas diffusion model was developed, revealing that the precision of the direct measurement is $60^{+35}_{-36}\times{10}^{-9}$~Pa$\cdot$ m$^3$/s/PPM. This method also allows for the identification of leak locations with a precision of a few millimeters. By utilizing a container outside the test object to accumulate leaked SF$_6$, overall leakage can be detected with a sensitivity better than ${10}^{-9}$~Pa$\cdot$~m$^3$/s; however, this sensitivity is primarily limited by the effectiveness of the container's sealing.

A dedicated leak test system was developed and applied to 203 electronics UWBs in the JUNO 3-inch PMTs system. A box accommodating the shape of the UWB was made playing the role of SF$_6$ accumulation. Calibration of the accumulation box was conducted and the leak rate $L$ was found to be smaller than $10^{-6}$~m$^3$/s. As a result, with the minimum accumulation time of 5.6 hours, leaks in UWBs were excluded at a rate of $2.4\times{10}^{-9}$~Pa$\cdot$ m$^3$/s in terms of SF$_6$, or $1.65\times{10}^{-8}$~Pa$\cdot$ m$^3$/s in terms of helium equivalent. This meets the sensitivity requirement of 1$\times$10$^{-7}$~Pa$\cdot$m$^3$/s. Based on our experiences, achieving a detection limit of $O(10^{-9})$~Pa$\cdot$m$^3$/s for SF$_6$ primarily relies on effectively sealing the accumulation box that encloses the test object. We find that butyl rubber adhesive tape serves as an excellent sealant, particularly due to its ease of use on uneven surfaces. To enhance sensitivity further, employing a higher-precision SF$_6$ detector or ensuring a more efficient seal of the accumulation box over an extended period is necessary.

Four UWBs were found to have leaks by the SF$_6$ accumulation system, and the leak points were identified by the SF$_6$ direct measurement. Before replacing the leaked components, two UWBs were also tested by a helium based system. Results of the three methods showed good consistency when taking into account uncertainties.

%The electronics of 3-inch PMTs were put in stainless steel under-water boxes (UWBs) and installed below water surface deepest about 40~m. Helium was firstly chosen as trace gas to do leak test, but helium may damage PMTs, and we chose SF$_6$ as trace gas finally. 

%Both direct measurement and accumulation methods were used to do leak test. For direct measurement method, calculation shows a sensitivity of $60^{+35}_{-36}\times{10}^{-9}$~Pa$\cdot$ m$^3$/s/PPM(need update), which is consistent with the empirical formula provided by the manufacturer of the SF$_6$ detector, which is $50\times{10}^{-9}$~Pa$\cdot$ m$^3$/s/PPM. But direct measurement is not stable since distances between sniffer and leak point and convection is uncontrollable. For accumulation method, we designed and made two transparent PC boxes to accumulate potential SF$_6$ leak into the PC boxes. Leak rate of PC boxes were calibrated five times and results show $L=5.3\pm4.7\times10^{-7}~m^3/s$. After accumulation for 6 hours, if no leak was found, the maximum leak rate of UWB is $2.3\times{10}^{-9}$~Pa$\cdot$ m$^3$/s.

%All 203 UWBs were tested with this leak test system. Some UWBs were found leak because miss assembling of O-rings and than fixed immediately. 4 UWBs were found leak and repaired by factory. Finally, all UWBs passed leak test. 200 UWBs were already installed to JUNO detector. Up to now, 200 UWBs are 

%2 leak UWBs were tested with helium mass spectrometer, and results were consistent with SF$_6$ results. 

The sensitivity of the leak test for the multiple O-rings system was explored through simulation. Due to the gradual accumulation of gas and the slow increase in pressure between the two O-rings, the overall leak rate is substantially lower than that observed in the single O-ring scenario. For the double O-ring in the receptacle, the detection limit is approximately $O(10^{-8})$~Pa$\cdot$m$^3$/s for SF$_6$ given our standard leak test duration.
%We discussed cascade leak situation to get true leak rate of double O-rings. Gas leak from the first O-ring will accumulate at the space between two O-rings and than leak out the second O-ring. This process will extend time needed to find potential leak point and we will not be able to find the issue if only one O-ring is invalid. 

Finally, all 200 UWBs were installed in the JUNO detector by the end of 2024, distributed approximately uniformly within a depth range of 2.7 meters to 40.6 meters in the water pool. Since the completion of water filling on February 1, 2025, and until the time of publication, a preliminary analysis shows that all 200 electronic components in the UWBs have been functional, indicating that there has been no water penetration. The final conclusion will be reported by the JUNO collaboration later.% Check with JPC

\backmatter
\section*{Declarations}
\bmhead{Acknowledgements}
Supported by the Strategic Priority Research Program of the Chinese Academy of Sciences, Grant No. XDA10011200.
We are grateful to the JUNO central detector group and the PMT instrumentation group for their recommendation of SF$_6$ as the tracer gas and for providing the SF$_6$ detector. We also extend our thanks to the JUNO liquid scintillator group for supplying the helium mass spectrometer and useful discussions.
\begin{CJK}{UTF8}{gbsn}
\bibliography{sn-bibliography}% common bib file


\begin{thebibliography}{17}
\ifx \bisbn   \undefined \def \bisbn  #1{ISBN #1}\fi
\ifx \binits  \undefined \def \binits#1{#1}\fi
\ifx \bauthor  \undefined \def \bauthor#1{#1}\fi
\ifx \batitle  \undefined \def \batitle#1{#1}\fi
\ifx \bjtitle  \undefined \def \bjtitle#1{#1}\fi
\ifx \bvolume  \undefined \def \bvolume#1{\textbf{#1}}\fi
\ifx \byear  \undefined \def \byear#1{#1}\fi
\ifx \bissue  \undefined \def \bissue#1{#1}\fi
\ifx \bfpage  \undefined \def \bfpage#1{#1}\fi
\ifx \blpage  \undefined \def \blpage #1{#1}\fi
\ifx \burl  \undefined \def \burl#1{\textsf{#1}}\fi
\ifx \doiurl  \undefined \def \doiurl#1{\url{https://doi.org/#1}}\fi
\ifx \betal  \undefined \def \betal{\textit{et al.}}\fi
\ifx \binstitute  \undefined \def \binstitute#1{#1}\fi
\ifx \binstitutionaled  \undefined \def \binstitutionaled#1{#1}\fi
\ifx \bctitle  \undefined \def \bctitle#1{#1}\fi
\ifx \beditor  \undefined \def \beditor#1{#1}\fi
\ifx \bpublisher  \undefined \def \bpublisher#1{#1}\fi
\ifx \bbtitle  \undefined \def \bbtitle#1{#1}\fi
\ifx \bedition  \undefined \def \bedition#1{#1}\fi
\ifx \bseriesno  \undefined \def \bseriesno#1{#1}\fi
\ifx \blocation  \undefined \def \blocation#1{#1}\fi
\ifx \bsertitle  \undefined \def \bsertitle#1{#1}\fi
\ifx \bsnm \undefined \def \bsnm#1{#1}\fi
\ifx \bsuffix \undefined \def \bsuffix#1{#1}\fi
\ifx \bparticle \undefined \def \bparticle#1{#1}\fi
\ifx \barticle \undefined \def \barticle#1{#1}\fi
\bibcommenthead
\ifx \bconfdate \undefined \def \bconfdate #1{#1}\fi
\ifx \botherref \undefined \def \botherref #1{#1}\fi
\ifx \url \undefined \def \url#1{\textsf{#1}}\fi
\ifx \bchapter \undefined \def \bchapter#1{#1}\fi
\ifx \bbook \undefined \def \bbook#1{#1}\fi
\ifx \bcomment \undefined \def \bcomment#1{#1}\fi
\ifx \oauthor \undefined \def \oauthor#1{#1}\fi
\ifx \citeauthoryear \undefined \def \citeauthoryear#1{#1}\fi
\ifx \endbibitem  \undefined \def \endbibitem {}\fi
\ifx \bconflocation  \undefined \def \bconflocation#1{#1}\fi
\ifx \arxivurl  \undefined \def \arxivurl#1{\textsf{#1}}\fi
\csname PreBibitemsHook\endcsname

%%% 1
\bibitem[\protect\citeauthoryear{Abusleme et~al.}{2022}]{JUNO:2022hxd}
\begin{barticle}
\bauthor{\bsnm{Abusleme}, \binits{A.}}, \betal:
\batitle{{JUNO physics and detector}}.
\bjtitle{Prog. Part. Nucl. Phys.}
\bvolume{123},
\bfpage{103927}
(\byear{2022})
\doiurl{10.1016/j.ppnp.2021.103927}
\end{barticle}
\endbibitem

%%% 2
\bibitem[\protect\citeauthoryear{Abusleme et~al.}{2021}]{JUNO:2020xtj}
\begin{barticle}
\bauthor{\bsnm{Abusleme}, \binits{A.}}, \betal:
\batitle{{Calibration Strategy of the JUNO Experiment}}.
\bjtitle{JHEP}
\bvolume{03},
\bfpage{004}
(\byear{2021})
\doiurl{10.1007/JHEP03(2021)004}
{\href{https://arxiv.org/abs/2011.06405}{{arXiv:2011.06405}}}
{[physics.ins-det]}
\end{barticle}
\endbibitem

%%% 3
\bibitem[\protect\citeauthoryear{Cabrera et~al.}{2024}]{Cabrera:2023dek}
\begin{barticle}
\bauthor{\bsnm{Cabrera}, \binits{A.}}, \betal:
\batitle{{Multi-calorimetry in light-based neutrino detectors}}.
\bjtitle{JHEP}
\bvolume{12},
\bfpage{002}
(\byear{2024})
\doiurl{10.1007/JHEP12(2024)002}
{\href{https://arxiv.org/abs/2312.12991}{{arXiv:2312.12991}}}
{[hep-ex]}
\end{barticle}
\endbibitem

%%% 4
\bibitem[\protect\citeauthoryear{Abusleme et~al.}{2023}]{JUNO:2022qgr}
\begin{barticle}
\bauthor{\bsnm{Abusleme}, \binits{A.}}, \betal:
\batitle{{JUNO Sensitivity on Proton Decay $p\to \bar\nu K^+$ Searches}}.
\bjtitle{Chin. Phys. C}
\bvolume{47}(\bissue{11}),
\bfpage{113002}
(\byear{2023})
\doiurl{10.1088/1674-1137/ace9c6}
{\href{https://arxiv.org/abs/2212.08502}{{arXiv:2212.08502}}}
{[hep-ex]}
\end{barticle}
\endbibitem

%%% 5
\bibitem[\protect\citeauthoryear{Zhang et~al.}{2025}]{Zhang:2024okq}
\begin{barticle}
\bauthor{\bsnm{Zhang}, \binits{S.-Y.}},
\bauthor{\bsnm{Huang}, \binits{Y.-B.}},
\bauthor{\bsnm{He}, \binits{M.}},
\bauthor{\bsnm{Yang}, \binits{C.-F.}},
\bauthor{\bsnm{Chen}, \binits{G.-M.}}:
\batitle{{Sub-GeV events energy reconstruction with 3-inch PMTs in JUNO}}.
\bjtitle{Nucl. Sci. Tech.}
\bvolume{36}(\bissue{5}),
\bfpage{84}
(\byear{2025})
\doiurl{10.1007/s41365-025-01678-4}
{\href{https://arxiv.org/abs/2402.13267}{{arXiv:2402.13267}}}
{[physics.ins-det]}
\end{barticle}
\endbibitem

%%% 6
\bibitem[\protect\citeauthoryear{Abusleme et~al.}{2025}]{JUNO:2024fdc}
\begin{barticle}
\bauthor{\bsnm{Abusleme}, \binits{A.}}, \betal:
\batitle{{Prediction of Energy Resolution in the JUNO Experiment}}.
\bjtitle{Chin. Phys. C}
\bvolume{49}(\bissue{1}),
\bfpage{013003}
(\byear{2025})
\doiurl{10.1088/1674-1137/ad83aa}
{\href{https://arxiv.org/abs/2405.17860}{{arXiv:2405.17860}}}
{[hep-ex]}
\end{barticle}
\endbibitem

%%% 7
\bibitem[\protect\citeauthoryear{Abusleme et~al.}{2022}]{JUNO:2022mxj}
\begin{barticle}
\bauthor{\bsnm{Abusleme}, \binits{A.}}, \betal:
\batitle{{Sub-percent precision measurement of neutrino oscillation parameters
  with JUNO}}.
\bjtitle{Chin. Phys. C}
\bvolume{46}(\bissue{12}),
\bfpage{123001}
(\byear{2022})
\doiurl{10.1088/1674-1137/ac8bc9}
{\href{https://arxiv.org/abs/2204.13249}{{arXiv:2204.13249}}}
{[hep-ex]}
\end{barticle}
\endbibitem

%%% 8
\bibitem[\protect\citeauthoryear{达道安}{2004}]{VaccumDesign}
\begin{bbook}
\bauthor{\bsnm{达道安}}:
\bbtitle{真空设计手册}.
\bpublisher{国防工业出版社},
\blocation{北京}
(\byear{2004})
\end{bbook}
\endbibitem

%%% 9
\bibitem[\protect\citeauthoryear{国家机械工业联合会}{2005}]{GB/T11605-2005}
\begin{botherref}
\oauthor{\bsnm{国家机械工业联合会}}:
湿度测量方法.
\url{https://openstd.samr.gov.cn/bzgk/gb/newGbInfo?hcno=DC8617DA62FED5133E26025F3EE08446}
Accessed 2025-4-22
\end{botherref}
\endbibitem

%%% 10
\bibitem[\protect\citeauthoryear{Chen et~al.}{2013}]{Chen:2012ef}
\begin{barticle}
\bauthor{\bsnm{Chen}, \binits{X.}}, \betal:
\batitle{{Leakage Tests of the Stainless Steel Vessels of the Antineutrino
  Detectors in the Daya Bay Reactor Neutrino Experiment}}.
\bjtitle{Sci. China Technol. Sci.}
\bvolume{56}(\bissue{1}),
\bfpage{148}
(\byear{2013})
\doiurl{10.1007/s11431-012-5007-2}
{\href{https://arxiv.org/abs/1203.0346}{{arXiv:1203.0346}}}
{[physics.ins-det]}
\end{barticle}
\endbibitem

%%% 11
\bibitem[\protect\citeauthoryear{}{}]{HeliumDetector}
\begin{botherref}
\url{https://www.agilent.com.cn/cs/library/usermanuals/public/VS\%20Series\%20Component\%20Leak\%20Detector.pdf}
\end{botherref}
\endbibitem

%%% 12
\bibitem[\protect\citeauthoryear{Incandela et~al.}{1988}]{INCANDELA1988237}
\begin{barticle}
\bauthor{\bsnm{Incandela}, \binits{J.R.}},
\bauthor{\bsnm{Ahlen}, \binits{S.P.}},
\bauthor{\bsnm{Beatty}, \binits{J.}},
\bauthor{\bsnm{Ciocio}, \binits{A.}},
\bauthor{\bsnm{Felcini}, \binits{M.}},
\bauthor{\bsnm{Ficenec}, \binits{D.}},
\bauthor{\bsnm{Hazen}, \binits{E.}},
\bauthor{\bsnm{Levin}, \binits{D.}},
\bauthor{\bsnm{Marin}, \binits{A.}},
\bauthor{\bsnm{Stone}, \binits{J.L.}},
\bauthor{\bsnm{Sulak}, \binits{L.R.}},
\bauthor{\bsnm{Worstell}, \binits{W.}}:
\batitle{The performance of photomultipliers exposed to helium}.
\bjtitle{Nuclear Instruments and Methods in Physics Research Section A:
  Accelerators, Spectrometers, Detectors and Associated Equipment}
\bvolume{269}(\bissue{1}),
\bfpage{237}--\blpage{245}
(\byear{1988})
\doiurl{10.1016/0168-9002(88)90885-6}
\end{barticle}
\endbibitem

%%% 13
\bibitem[\protect\citeauthoryear{Norton}{1957}]{10.1063/1.1722570}
\begin{barticle}
\bauthor{\bsnm{Norton}, \binits{F.J.}}:
\batitle{{Permeation of Gases through Solids}}.
\bjtitle{Journal of Applied Physics}
\bvolume{28}(\bissue{1}),
\bfpage{34}--\blpage{39}
(\byear{1957})
\doiurl{10.1063/1.1722570}
{\href{https://arxiv.org/abs/https://pubs.aip.org/aip/jap/article-pdf/28/1/34/18316282/34\_1\_online.pdf}{{https://pubs.aip.org/aip/jap/article-pdf/28/1/34/18316282/34\_1\_online.pdf}}}
\end{barticle}
\endbibitem

%%% 14
\bibitem[\protect\citeauthoryear{}{}]{LF-300}
\begin{botherref}
\url{http://www.kstone.cc/product2/53.html}
\end{botherref}
\endbibitem

%%% 15
\bibitem[\protect\citeauthoryear{{EDWARD N. FULLER, PAUL D. SCHETTLER, J.
  CALVIN GIDDINGS}}{1966}]{1966551}
\begin{botherref}
\oauthor{\bsnm{{EDWARD N. FULLER, PAUL D. SCHETTLER, J. CALVIN GIDDINGS}}}:
{A new method for prediction of binary gas-phase diffusion coefficients}.
Industrial and Engineering Chemistry
\textbf{58}(5)
(1966)
\end{botherref}
\endbibitem

%%% 16
\bibitem[\protect\citeauthoryear{Zarkova and Hohm}{2002}]{10.1063/1.1433462}
\begin{barticle}
\bauthor{\bsnm{Zarkova}, \binits{L.}},
\bauthor{\bsnm{Hohm}, \binits{U.}}:
\batitle{pvt–second virial coefficients b(t), viscosity η(t), and
  self-diffusion ρd(t) of the gases: Bf3, cf4, sif4, ccl4, sicl4, sf6, mof6,
  wf6, uf6, c(ch3)4, and si(ch3)4 determined by means of an isotropic
  temperature-dependent potential}.
\bjtitle{Journal of Physical and Chemical Reference Data}
\bvolume{31}(\bissue{1}),
\bfpage{183}--\blpage{216}
(\byear{2002})
\doiurl{10.1063/1.1433462}
{\href{https://arxiv.org/abs/https://pubs.aip.org/aip/jpr/article-pdf/31/1/183/8183598/183\_1\_online.pdf}{{https://pubs.aip.org/aip/jpr/article-pdf/31/1/183/8183598/183\_1\_online.pdf}}}
\end{barticle}
\endbibitem

%%% 17
\bibitem[\protect\citeauthoryear{Liu et~al.}{2025}]{LIU2025105517}
\begin{barticle}
\bauthor{\bsnm{Liu}, \binits{W.}},
\bauthor{\bsnm{Shi}, \binits{J.}},
\bauthor{\bsnm{Liu}, \binits{Y.}},
\bauthor{\bsnm{Chen}, \binits{Y.}},
\bauthor{\bsnm{Wu}, \binits{P.}},
\bauthor{\bsnm{Hou}, \binits{K.}},
\bauthor{\bsnm{Li}, \binits{X.}},
\bauthor{\bsnm{Zhang}, \binits{Y.}},
\bauthor{\bsnm{He}, \binits{M.}}:
\batitle{Measurement of helium thermophysical properties and modification of
  the calculation models in the kta 3102.1 report}.
\bjtitle{Progress in Nuclear Energy}
\bvolume{178},
\bfpage{105517}
(\byear{2025})
\doiurl{10.1016/j.pnucene.2024.105517}
\end{barticle}
\endbibitem

\end{thebibliography}
\end{CJK}

\end{document}